\documentclass[aps,a4paper,12pt,reprint,twocolumns,superscriptaddress]{revtex4-1}

\usepackage[T1]{fontenc}
\usepackage{graphicx}	
\usepackage{amsmath}	
\usepackage{amssymb}	
\usepackage{mhchem}
\usepackage{color}
\usepackage{longtable}

\newcommand{\jpca}{J.~Phys.~Chem.~A} 
\newcommand{\apjs}{Astrophys. J. Suppl. Serie.}
\newcommand{\apjl}{Astrophys. J. Letters}
\newcommand{\aap}{Astronomy \& Astrophysics}
\newcommand{\mnras}{Monthly Notices of the Royal Astronomical Society}
\newcommand{\ssr}{Space Science Review}
\newcommand{\physrep}{Physics Reports}
\newcommand{\jqsrt}{J. Quantitative Spectroscopy and Radiative Transfer}

\newcommand{ \red }[1]{{{{\color[rgb]{0,0,0}#1}}}} 

\begin{document}
\title{\textrm{\large Report from the Workshop on}\\ \Large
Theory 
of Gas Phase Scattering and
Reactivity for Astrochemistry\\
\textrm{\large Nov 23 -- Dec 4, 2015}}
\affiliation{IPAG, CNRS \& Universit\'{e} Grenoble-Alpes, France}\affiliation{CAS@MPE, MPE, Garching, Germany}
\affiliation{U. Algarve, Faro, Portugal}
\affiliation{Physique des Lasers Atomes et Mol\'ecules, Univ.de Lille \&  CNRS, Lille, France}
\affiliation{U. Virginia, USA}\affiliation{Chemical Sciences and Engineering Division
	Argonne National Laboratory, Argonne, IL, 60439 USA}
\affiliation{U. Sapporo, Japan}\affiliation{Women U. Nara, Japan}\affiliation{U. Bologna, Italy}\affiliation{CSIC, 
Madrid, Spain}
\affiliation{U. Nijmegen, The Netherlands}\affiliation{U. Groningen, The Netherlands}

\author{Laurent Wiesenfeld}\thanks{Editor, SOC member}\affiliation{IPAG, CNRS \& Universit\'{e} Grenoble-Alpes, 
France}

\author{Wing-Fai Thi}\thanks{Editor}\affiliation{CAS@MPE, MPE, Garching, Germany}

\author{Paola Caselli}\thanks{Editor, SOC member}\affiliation{CAS@MPE, MPE, Garching, Germany}

\author{Alexandre Faure}\thanks{Editor}\affiliation{IPAG, CNRS \& Universit\'{e} Grenoble-Alpes, France}

\author{Luca Bizzocchi}\thanks{Contributor}\affiliation{CAS@MPE, MPE, Garching, Germany}

\author{Jo\~ao Brand\~ao}\thanks{Contributor}\affiliation{U. Algarve, Faro, Portugal}

\author{Denis Duflot}\thanks{Contributor}\affiliation{Physique des Lasers Atomes et Mol\'ecules, Univ.de Lille \&  
CNRS, Lille, France}

\author{Eric Herbst}\thanks{Contributor}\thanks{SOC member}\affiliation{U. Virginia, USA}

\author{Stephen J. Klippenstein}\thanks{Contributor}\affiliation{Chemical Sciences and Engineering Division
	Argonne National Laboratory, Argonne, IL, 60439 USA}

\author{Tamiki Komatsuzaki}\thanks{Contributor}\thanks{SOC member}\affiliation{U. Sapporo, Japan}

\author{Cristina Puzzarini}\thanks{Contributor}\affiliation{U. Bologna, Italy}

\author{Octavio Roncero}\thanks{Contributor}\affiliation{CSIC, Madrid, Spain}

\author{Hiroshi Teramoto}\thanks{Contributor}\affiliation{U. Sapporo, Japan}

\author{Mikito Toda}\thanks{Contributor}\affiliation{Women U. Nara, Japan}

\author{Ad van der Avoird}\thanks{Contributor}\thanks{SOC member}\affiliation{U. Nijmegen, The Netherlands}

\author{Holger Waalkens}\thanks{Contributor}\affiliation{U. Groningen, The Netherlands}

\begin{abstract}
Because of the very peculiar conditions of chemistry in many 
 astrophysical gases (low densities, mostly low temperatures, kinetics-dominated chemical evolution), great 
efforts have been devoted to study molecular signatures and chemical evolution. While experiments are being 
performed in many 
laboratories, it appears that the efforts directed towards  theoretical works are not as strong. 

This report deals with the present status of chemical physics/physical chemistry theory, for the qualitative and 
quantitative understanding of kinetics of molecular scattering, being it reactive or inelastic. By gathering 
several 
types of expertise, from applied mathematics to physical chemistry,  dialog is made possible, as a step 
towards new and more 
adapted theoretical frameworks, capable of meeting the theoretical, methodological and numerical challenges of 
kinetics-dominated gas phase chemistry in  
astrophysical 
environments.  

A state of the art panorama is presented, alongside present-day strengths and shortcomings. However, coverage is 
not  complete, being limited in this report to actual attendance of the workshop. Some paths towards 
relevant 
progress are proposed.

\date{\today}

\vspace{4em}

\par\noindent
\textsf{\normalsize This workshop was organised at the initiative of the \textsf{COST Action CM1401} "Our 
Astrochemical History"~\footnote{PI, Laurent 
Wiesenfeld, \texttt{http://cost.obs.ujf-grenoble.fr/}}, with the generous support of the \textsf{CAS@MPE} laboratory\footnote{Dir., Paola Caselli, 
\texttt{http://www.mpe.mpg.de/CAS} },
 of the \textsf{MIAPP}\footnote{\texttt{http://www.munich-iapp.de}}, and  the \textsf{CNRS}
\footnote{\texttt{http://www.cnrs.fr/mi/spip.php?article621\&lang=fr}}.
 }

\end{abstract}
\maketitle
\clearpage
\newpage

\normalsize

\section{Synthetic presentation}
We begin with a very broad presentation, and setting up the scene of chemistry.
Next, we try and make an analytical description of the work in MIAPP/Garching. We finish with perspectives, and a 
table 
of the participants.

\subsection{Gas phase chemistry in Astrophysics} 
 Chemistry in interstellar environments has many peculiarities, that makes it very different from chemistry in 
environments that are customary for the laboratory 
chemist. While the basic concepts of physical chemistry evidently remain, tLhe physical conditions 
are 
such that many of the usual hypotheses do not hold.
Being it interstellar gaseous matter (ISM),  gases surrounding  Solar System 
objects like comets and asteroids, the main constraints of the kind of 
chemistry we are dealing with are: (i) Very dilute conditions, number density $\rho \lesssim 10^{10}\,
\mathrm{cm^{-3}}$
\;
\footnote{Under normal conditions, 298K, 1 atmosphere, ideal gas has a density of $2.5\,10^{19}\,\mathrm{cm^{-3}}
$}; 
(ii) 
Large range of temperatures, with molecules 
observed at gas temperatures as low as 5-15~K; (iii) Overwhelming dominance of atomic and molecular hydrogen, 
which 
constitute more that 90\% 
(in number) of the whole 
number density, all other elements but  inert He being less than 1\% in number. Main elements are   O, C, Ne, Fe, 
N, 
Si, Ar, Mg, S, and all others  present as traces, including alkaline elements, halogens and P.

As a consequence of the physical and chemical conditions, the main differences between laboratory chemistry and 
astrochemistry, which are relevant for this workshop could be described as follows:
\begin{enumerate}
\item Because of the low density and temperature in the ISM, the chemistry is never at thermodynamical 
equilibrium. 
Chemical processes are dominated by the 
kinetics and branching ratios of the reactions at hand, and  are at  steady state at best, even if astrophysical timescales are relatively long, 
characteristic times 
being of the order of $10^4 - 10^6$ years. Only in planetary atmospheres   thermal equilibrium is reached.
\item In space, three-body collisions are absent (or extremely unlikely). This means that reactions like 
$A+B
 \rightarrow AB $ may only occur with  photonic stabilisation, or else on the surface of a grain.
 
\item The often prevailing low temperatures of interstellar space prevent many reactions to take place, even with modest activation 
barriers.  However, some reactions proceed
through tunneling, especially so if involving atomic H (or D).
\item Ion chemistry plays a important role in space, mainly through cations (though anions are also present). Main ions are 
atomic species (in diffuse clouds) and 
protonated species (in molecular clouds).
\item In interstellar dilute environments,  ionization of trace elements, such as C and S, may proceed by photon with $E < 13.6$~eV, photon of higher 
energy 
being 
absorbed by atomic H. However, cosmic rays (CR) are ubiquitous, even in the denser parts of the ISM and are the 
main 
primary source of energy that initiate chemical chains (by the reaction summarized by $ \mathrm{  H_2 + CR 
\rightarrow e^{-} 
+ H_2^+}$ followed by $\mathrm{ H_2+ H_2^+\rightarrow H_3^+ + H}$).
\item For many environments of very low ionisation fraction,
the neutral-neutral reactions are of importance, especially so those involving radicals tend to have higher 
rates at 
low temperatures \citep{Faure2009JPCA..11313694F}. However, ion/electron  chemistry dominates the chemical 
evolution in 
many 
sectors of molecular complexification, like the successive hydrogenation of \ce{N^+} or CH$^+$.
\end{enumerate}
Great details may be found in many reviews (see e.g. \cite{Dalgarno:2008aa,Herbst:2009aa,Smith:2011aa,
Caselli:2012aa}). 
Let us also underline that the chemistry of the ISM is relatively rich, with almost two hundred molecules 
detected, 
some 
of them quite unusual on the point of view of the laboratory: Long carbon chains, possibly charged --anions and 
cations--, and substituted (CH$_3$, CN); some  cyclic molecules (\ce{c-C3H2}), many radicals and protonated species, like the common 
\ce{HCO^+}, \ce{N_2H^+}, \ce{H_2O^+} and the all-important \ce{H_3^+}.

The precise 
identification of a chemical species (including all its isotopologues) is made possible from observed rotational/spectra from cm to FIR wavelengths and thanks to the fundamental spectroscopic experimental work which provide accurate frequencies to compare with.
It makes use of large spectroscopic databases, still not precise enough especially so in the frequency region 
above 
800--1000~GHz. Also, many isotopologues of the main molecules are detected, with D, \ce{^{13}C}, \ce{^{15}N}, 
\ce{^{17}
O}, 
\ce{^{18}O} as main substitutes; even some multiply substituted isotopologues are detectable, like \ce{^{13}C^{18}O} or \ce{D^{13}CO^{+}}. 
Complete spectroscopic databases for isotopologues, and for vibrational excited states are far from complete, and 
computation of rotational lines is still beyond possibilities, at the precision required ($\Delta\nu/\nu \leq 
10^{-6}$).
Present databases include the CDMS in Cologne  \citep{Muller:2005bh} and the JPL in Pasadena \citep{drouin:JPL}  
with 
other data bases 
found in links in the CDMS front page.

The quantitative description of the ISM necessitates a non equilibrium model of the radiative transfer of the most 
abundant molecular species. In order to achieve this, it has been known since a long time that the excitation/de-excitation 
of 
the 
various molecular lines is determined both by photon emission/absorption and by collisions with the main components of the 
molecular gas, or with electrons. In the ISM, the projectiles are mainly \ce{H_2}, H, \ce{e^-}. For planetary atmospheres, dense regions of protoplanetary disks and atmospheres of Solar System objects, the main gas may be heavier, \ce{H_2O}, \ce{CO_2}, \ce{N_2}, or \ce{CH_4}. In any case, the rates of 
energy exchange between the projectile and the target --the molecule being observed-- determine the lines 
intensities 
for 
most rotational transitions, and also for some ro-vibrational transitions, if the FIR emission rate is low enough. 
This 
is 
especially true for electric-dipolar forbidden transitions, like magnetic  (fine-structure) transitions in open 
shell 
atoms or radicals. Also, elastic/inelastic scattering computation open the way to ab 
initio 
computing of pressure broadening  \citep{Drouin2012PhRvA..86b2705D} and other collisional properties
 \citep{Song:2015zr,Karman:2015kx}.

Most of those statements have been known for years, and the importance of microscopic knowledge is fully 
recognized 
nowadays (see e.g. the special issue of Chemical Reviews, 2013, Volume 113, Issue 12). It has been a major 
endeavour, to  
try and model 
 rates of reactions and incorporate them in the chemical reaction networks. Widely used networks may be 
comprehensive 
and 
even include surface 
reactions and adsorption/desorption \citep{Wakelam:2012ul,Wakelam:2015qf,McElroy:2013ly}. Other networks are 
specialised 
for the chemical conditions of 
interest, like PDR's or very cold pre-stellar core. Modern simulation tools include chemical models coupled to 
physical dynamics, in particular hydro- and magneto-hydro-dynamics. Change of thermal or transport properties because chemistry may occur and influence back hydrodynamics of the considered medium. Any 
of those networks necessitates the knowledge of $10^3-10^($ reaction rates, at low pressure, and at 
variable 
temperatures and pressures. 

It must not be forgotten that the characteristics of astrochemistry presented so far make it certainly peculiar, but 
it shares some of those peculiarities with a few other branches of physical chemistry. Let us mention here two representative cases:
\begin{itemize}
\item
The diluted gas chemistry is prevalent in upper atmospheres, and ours too. Chemistry in our upper atmosphere, 
while 
quite 
distinct from the ISM (importance of the Solar wind and photons, massive presence of O, Cl, role of aerosols),  presents
some 
mechanisms and conditions that are quite similar. 
\item
Also, for many years, combustion chemistry has been dealing with various radicals, built with the H, C, N, O 
elements 
principally. It is no chance that many of the chemical rates used in astrochemical databases originate in one way 
or 
another from rates used in combustion chemistry, even if physical conditions may be vastly 
different, see the massive database~\cite{Baulch:2005aa}.
\end{itemize}
\subsection{Aim of this workshop}

Because of these very specific conditions, it has been recognized that much experience has to be gained, in order 
to 
understand model reliably the chemical evolution of matter in astrophysical environments. Several large 
agencies and 
several European and national initiatives tend to fund and favor 
research in the domain of \emph{Laboratory Astrochemistry}, as it is presently called. While  many experiments are 
running or are
being set up in numerous laboratories, it has become clear lately that the chemical physics/physical chemistry 
theory
that would adequately model astrochemistry is  neither at the level of the astrophysics models, nor  as 
sophisticated as 
the experiments which are now conceived and put into operations. 

\textbf{
We  feel it useful to describe where theory stands now in several European and international teams. We choose to 
focus on one specific subject, because of the width of possible themes: Gas phase molecular
 scattering, whether inelastic or reactive. 
}

We have the aim to compute rates of energy exchange, rates and branching ratios of chemical reactions, in the 
temperature, pressure and abundance ranges relevant for astrophysical problems. We wish to have mathematical and 
physical theories, 
computer codes and models that are able to produce reliable, quantitative models of those astrophysics objects 
which are 
rich in molecules interacting with each other.

\begin{figure*}[htb]
\begin{center}
\includegraphics[width=0.70\textwidth]{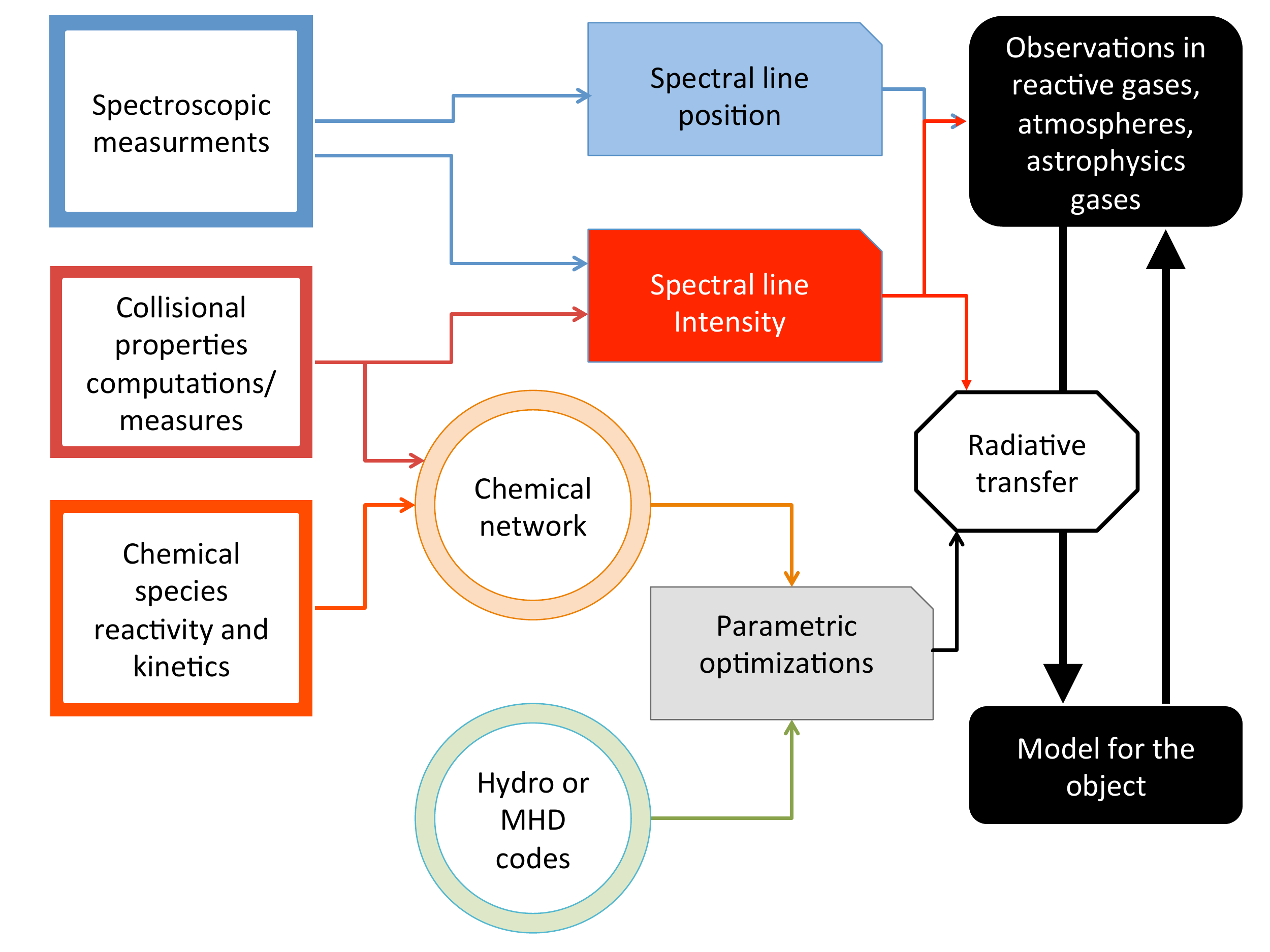}

\caption{Schematic flow of information from the observation to the model of the objects, with inputs of molecular 
physics: spectroscopy, collisions, reactivity.}
\label{fig:scheme1_1}
\end{center}
\end{figure*}

Two diagrams are presented below (Figures~\ref{fig:scheme1_1}-\ref{fig:scheme1_1}), which illustrate the goals and aims of this workshop. On both diagrams, 
the main themes of this workshop are the square boxes in red, and spectroscopy are the square blue 
boxes.

In the first one, one sees 
the usual way of describing the flow of information, back and forth from the observation on the sky to an 
understanding 
of the object, thanks to model presenting a spatial (or at least radial) structure in density, temperature, gas 
and 
grain 
content, degree of turbulence. The scheme underlines the necessity of the several levels of simulation that concur 
into 
finding a good model, validated by the conformity with observation, of  lines shapes and intensities of the 
molecular 
species as well as the continuum maps or other maps of the object in the sky.

The second diagram shows the various inputs towards chemical physics/physical chemistry relevant properties, 
emphasizing 
both experimental and theoretical inputs.

Let us underline the importance of observation-
induced 
molecular data (including species, like \ce{c-C_3H_2}, or some anion whose spectroscopy used to be unknown, like 
\ce{C_6H^-}, 
\cite{McCarthy2006ApJ...652L.141M}). It is instructive to note that \ce{CH^+}, even if known in principle, was 
first 
detected in 
space by its optical transitions, then found in experimental plasmas by\citet{Douglas1941ApJ....94..381D}, in the 
very 
beginnings 
of molecular physics for astrophysics.
 
\begin{figure*}[hbt]
\begin{center}
\includegraphics[width=0.70\textwidth]{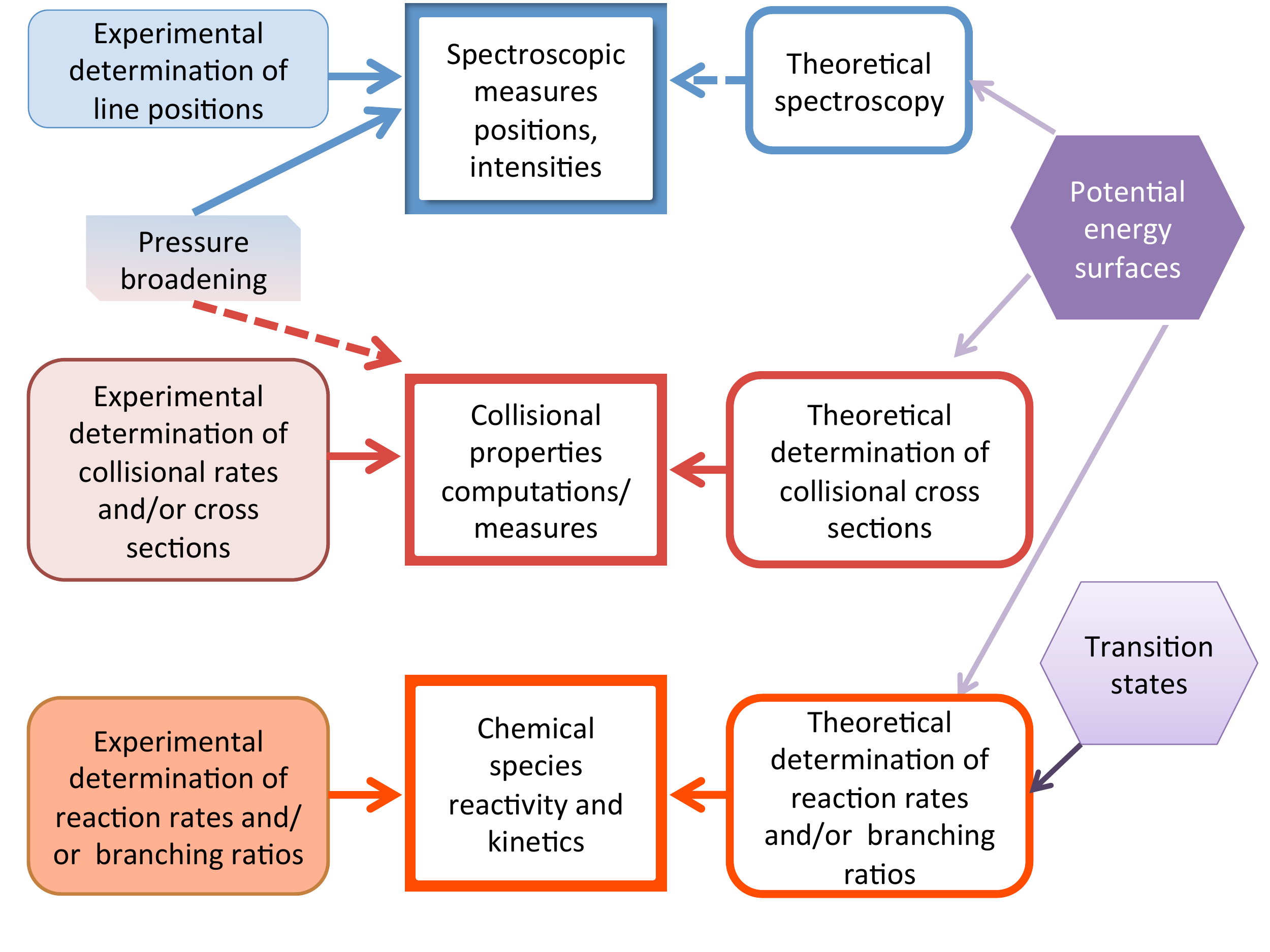}
\caption{Schematic ways to determine the dynamical and  spectroscopic data needed for figure~\protect\ref{fig:scheme1_1}. Experiments on 
the 
left,  theory on the right. Note  the 2 sources of dynamics, either the full PES surface or the stationary points 
of the 
Transition States.}
\label{fig:scheme1_2}
\end{center}
\end{figure*}

The goal of this worksop, on a synthetic/diagrammatic view, is
 to list (at least partially) the theoretical methods that are currently in use, for dealing with  the 
\emph{theoretical 
part} of the program, in the right hand side of figure~\ref{fig:scheme1_2}. The theoretical goals split into 
two parts; the 
static part (or electronic part) of the program consists in determining the landscape in which the atoms (or 
nuclei) 
move 
(violet hexagons in figure~\ref{fig:scheme1_2}, and also section \ref{sec:PES}). The dynamics part, in reddish in figure~\ref{fig:scheme1_2}, deals 
with 
the much more complex problem of how do these atoms move in the landscape,including making and destroying 
chemical 
bounds and/or undergoing non adiabatic transitions (transfers of spin, electrons, protons).
  
%
%
\clearpage
\section{Analytical presentation}\label{sec:analy} 
In the present chapter~\ref{sec:analy}, sections are written by the author(s) explicitly stated. If none stated, 
section is written by the editors (see the footnotes in the list of authors).
\subsection{ Chemical Simulations of Interstellar Sources}

\noindent
Text by \textsc{ E. Herbst}\\

\begin{figure*}[htbp]
\begin{center}
\includegraphics[width=0.75\textwidth]{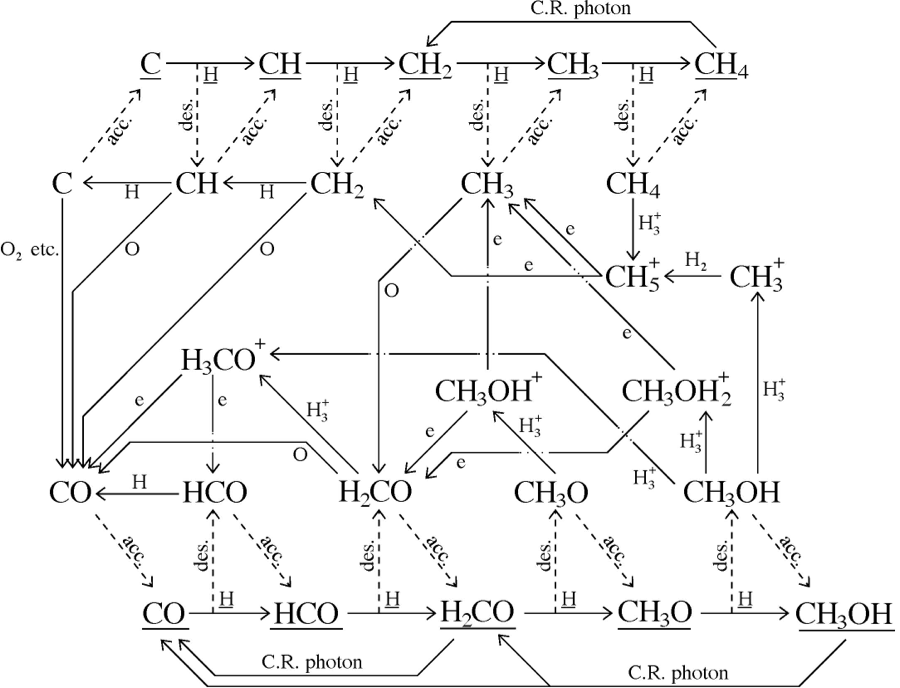}
\caption{%
Low temperature chemical routes involving the conversion among solid and gaseous methanol, carbon dioxide, carbon 
monoxide, and methane.  Grain-surface species are underlined. 'acc' and 'des' refer to adsorption onto grains and 
desorption from grains. Figure taken from \cite{Garrod2007A&A...467.1103G}}
\label{fig:eh}
\end{center}
\end{figure*}

The simulation of the chemistry that occurs in sources of the interstellar medium (ISM), in particular the dense 
interstellar medium, requires a large number of reactions, both in the gas phase and on the surfaces of dust 
particles 
\cite{Wakelam2010SSRv..156...13W}.  Networks of chemical reactions now range in size from 10,000 to over 100,000 
reactions depending 
upon the number of isotopomers included, especially those that include deuterium 
\cite{Albertsson2013ApJS..207...27A}.  
The newest 
networks are assumed to be useful at temperatures from 10 K to 800 K, but only a small fraction of the reactions 
have 
rate coefficients studied over such a wide range of temperature, often necessitating extrapolation or additional 
theoretical or experimental effort based on sensitivity analyses \cite{Acharyya2015MolPh.113.2243A}.   Most of the 
gas-
phase 
reactions in the networks are ion-neutral processes, with the ions produced by bombardment of cosmic rays, mainly 
high-
energy protons \cite{Herbst1973ApJ...185..505H}.  Reactions involving neutral species can also be important, 
despite the 
low temperatures in most of the ISM \cite{Balucani2015MNRAS.449L..16B}. In addition, there are dissociative 
recombination reactions, 
in which molecular positive ions combine with electrons to form neutral fragments, neutral-neutral reactions that 
occur 
via tunneling \cite{Acharyya2015MolPh.113.2243A} and radiative association reactions, in which a complex is 
formed, 
which then 
stabilizes by emission of a photon \cite{Wakelam2010SSRv..156...13W,Wakelam:2015qf}.  These processes were designed initially to study cold 
dense 
clouds with typical temperatures of 10 K.  For higher temperature sources, in which star formation has proceeded 
at 
least 
partially, neutral-neutral reactions with small barriers including endothermic processes assume more importance 
and have 
been subsequently included \cite{Harada2010ApJ...721.1570H}. An example of network is presented in figure \ref{fig:eh}.

The reactions that occur on dust particle surfaces are assumed for the most part to occur via the Langmuir-
Hinshelwood 
mechanism, which is based on diffusive motions of the surface reactants until they collide with one another and 
react 
\cite{Herbst2014PCCP...16.3344H}.  This diffusive motion depends on molecular vibrational frequencies in potential 
wells 
coupled with 
tunneling or hopping of species from one well to another.  Often, the barriers against hopping and tunneling are 
not 
known accurately for adsorbate-substrate pairs.  The gas-phase and grain-surface chemistries are coupled by the 
processes 
of adsorption from the gas onto the grains, and desorption from the grain surfaces back to the gas.  Adsorption is 
most 
efficient at the lowest temperatures, when the adsorbate sticks to the surface with only weak van der Waals-like 
binding, 
known as physisorption.  Thermal desorption, or sublimation,  is inefficient at low interstellar temperatures (10 
-- 20 
K) 
except for hydrogen and helium \cite{Hasegawa1992ApJS...82..167H}.  Non-thermal desorption at these temperatures 
occurs 
via a variety 
of mechanisms, mainly photodesorption and reactive desorption 
\cite{Bertin2016ApJ...817L..12B,Minissale2016MNRAS.458.2953M}; the former has 
been studied in detail in recent years, while the latter has received much less study.  The mechanism involves the 
transfer of a portion of the exothermicity of a surface reaction into kinetic energy able to eject the product 
from the 
surface.

To solve the kinetics so as to determine abundances, one can use rate equations for each species in the network, 
and 
integrate these coupled differential equations as a function of time \cite{Hasegawa1992ApJS...82..167H}.  The 
computer 
coding for 
this process is referred to as a 'chemical model' and the result a  'chemical simulation'. The steady-state 
condition is 
rarely reached before unphysically long times.  Instead, large mantles of ices, mainly water, carbon dioxide, and  
carbon 
monoxide,  grow in colder regions.  Rate equations exist for both gas-phase and grain-surface species although the 
rate 
coefficients for the two phases are quite different, the former governed by trajectories involving both short-
range and 
long-range forces, while the latter approximate a random, diffusive motion.  The simplest treatment with rate 
equations 
is known as the two-phase approach, with the two phases referring to the gas and the surfaces of the dust 
particles.   
In 
other words, if an ice mantle of many monolayers exists, the chemistry occurring atop the highest monolayer is not 
differentiated from the chemistry underneath in the bulk ice.  An extremely different approximation is known as 
the 
three-phase treatment  \cite{Hasegawa1993MNRAS.263..589H,Garrod2013ApJ...765...60G}; here, it is most often 
assumed that 
the bulk of the 
ice mantle undergoes no reactive chemistry at all.   Both of these treatments are approximate for surface 
chemistry for 
a 
number of reasons.  First, the chemistry occurring in the bulk is likely slower than that occurring on the 
surface, but 
is hardly occurring at a zero rate \cite{Lauck2015ApJ...801..118L}.  Secondly, the use of rate equations, while 
perfectly fine for 
gas-phase processes, is only approximate for surface reactions, based on the small number of reactive species per 
grain.  
More advanced treatments than the solution of rate equations are based on discrete but stochastic approaches. 
These 
approaches yield the number of species per grain, the uncertainty in this number, and the positions where the 
species 
lie 
in each monolayer.  The most utilized stochastic approach today is a Monte Carlo realization of the stochastic 
nature of 
the physics and chemistry on a grain \cite{Garrod2013ApJ...778..158G,Chang2016ApJ...819..145C}. Unfortunately, the 
use 
of a microscopic 
Monte Carlo method to solve the grain kinetics necessitates use of a similar, but macroscopic procedure for the 
gas-
phase 
kinetics so that the clock for both chemistries is the same. Current computational methods for the Monte Carlo 
procedure are exceedingly 'expensive' and beset by some flaws.  Not surprisingly, they can only be used for 
astronomically short periods of time at low temperature.  Approximations for the Monte Carlo procedure can do 
somewhat 
better, but at the expense of limited accuracy \cite{Vasyunin2013ApJ...762...86V}.

Chemical simulations can be undertaken for a wide diversity of interstellar sources, many of which are 
evolutionary 
stages in the formation of stars and planets.  If the physical conditions change during the existence of a 
particular 
stage, and if they are also heterogeneous, then the computer time and memory needed for full solution of the 
physics and 
the chemistry can be quite extensive, even if only rate equations are utilized for the chemistry.  For example, 
the 
evolutionary stages of solar-type systems involve the gradual formation under the influence of gravity of so-
called 
'cold 
cores', which possess gas densities of  10$^4$ cm$^{-3}$, mainly H$_2$, and gas and grain temperatures of 10~K 
\cite{Herbst:2009aa}.
 The gas-phase molecules found in cold cores contain many exotic species by terrestrial standards; these 
include radicals, linear carbon chains, metastable isomers, isotopologues, cations, anions, and three-membered 
rings. 
Some cold cores collapse to dense objects known as pre-stellar cores, which possess higher densities in a central 
condensation where much of the gas phase heavier than H$_2$ is lost to mantles on grains.   Eventually, the system 
becomes 
adiabatic and starts to heat up as material collapses onto the central condensation, now known as a protostar.  As 
the 
collapsing gas and dust,  now labelled a hot core,  reach temperatures of 100 -- 300 K, the molecular inventory 
changes 
to 
a terrestrial-type chemistry, with common organic molecules \cite{Crockett2015ApJ...806..239C}.  It is currently 
thought 
that at 
least some of these species are formed on warm dust particles and then sublimate into the gas phase as the dust 
particles 
reach even higher temperatures \cite{Garrod2008ApJ...682..283G}.  In addition to the hot core, a nearly planar and 
dense 
disk forms 
around the protostar, and dust particles in the disk coagulate to form comets, meteors, asteroids, and even 
planets.  
The chemistries of all evolutionary stages of low-mass and high-mass star formation have been studied, with some 
stages 
better known that others.  Protoplanetary disks are of great interest, but are small objects difficult to study 
observationally until recently, when a new generation of telescopes came into use.  Most recently, the organic 
molecule 
methanol was detected in the gas phase of a protoplanetary disk, in partial agreement with a prior chemical 
simulation 
\cite{Walsh2016ApJ...823L..10W}.  Advanced chemical simulations of protoplanetary disks should be combined with 
three-
dimensional 
hydrodynamics to follow the collapse of a hot core into such a disk as the chemistry also occurs.  At the present 
stage, 
such calculations do not quite reach the physical conditions observed for protoplanetary disks
 \cite{Furuya2013ASPC..476..385F}.  
Nevertheless, the chemistry of planetary atmospheres surrounding exo-solar planets that are formed in older 
protoplanetary disks is a new field destined to achieve popularity.  

%
\subsection{At the root : Potential Energy Surface}\label{sec:PES}

\noindent
Text  by \textsc{Denis Duflot and Jo\~{a}o Brand\~{a}o}\\

In order to compute dynamical effects, there are many ways to take care of the forces between
the atoms and molecules, depending mainly on the size of the problem at hand the precision
desirable.

In all of the theoretical works described here, because of the reduced number of atoms, the concept of a complete 
Potential Energy Surface (PES) is valid and its usefulness is not challenged. Working on model surfaces was 
considered 
as 
non-relevant, except for some very mathematical works, or proofs of concept. This very 
strong 
statement stems from several concurring factors, which make the computation of PES the most reliable part of our 
program. 
It is absolutely obvious from all interventions that computing reactive or inelastic scattering dynamics 
necessitates a 
PES of 'high' quality, 'high' meaning here that the various approximations are carefully described and justified, 
see 
below. There seemed to be three reasons for that: (i) the constant development of more versatile and more precise 
formalisms in quantum chemistry; (ii) the existence of well-balanced suitable suites of gaussian atomic basis sets 
for 
the description of molecular wave functions; (iii) the rapid implementation of state-of-the-art equations in 
actual {\sl 
ab initio} codes, widely available, either as free codes or as or commercial ones, both thoroughly maintained and 
supported. One of the best example is the {\bf\tt MOLPRO} code, particularly well adapted to our needs. Those 
codes 
incorporate many high precision methods: Coupled Clusters even at a higher level 
\cite{Lane2013doi:10.1021/ct300832f}, 
the so-called golden 
standard CCSD(T) \cite{Bartlett2007RevModPhys.79.291}, explicitly correlated (F12) treatment of electronic cusps 
\cite{Hattig2012doi:10.1021/cr200168z,Kong2012doi:10.1021/cr200204r}, special methods 
for long range interactions, like SAPT \cite{Jeziorski1994doi:10.1021/cr00031a008,Misquitta2005JChPh.123u4103M}. A 
description of the {\bf\tt MOLPRO} code is found in \cite{MOLPRO,Werner2012WCMS:WCMS82}. Also, these codes are 
able to handle at different level of precision open and closed-shell 
problems, and even 
multi-configuration effects (several arrangements of the electrons among the available electronic levels), albeit 
in a 
less straightforward and less rapid way.
Computation times are in general not negligible, and the quantum chemistry codes are not fully parallelized 
(generally 
limited to one node, to avoid inter-node communication). However, because the description of the PES necessitates 
computing usually a large number $N$ of points, trivial parallelization is usually possible. It  actually scales 
in a 
not 
straightforward way with the number of degrees of freedom, thanks to clever Monte-Carlo explorations or else to 
good 
fitting schemes (see below). This is feasible for small systems, but for larger systems strategies are needed to 
reduce 
the calculation to the chemical important region. As the computation of each point increases with the fourth power 
of 
number of basis set functions, stemming from the dimension of the matrices and the number configurations used in 
the 
calculation, the accurate energy estimation of each point became a crucial task. 
Besides, the main bottlenecks reside often, on the one hand, in the proper tests of the {\sl ab initio} methods 
(choice 
of basis set, calibration), and on the other hand, in the fitting procedures. 

In practice, experience is that 1 ab initio point/cpu is not the best choice. The number of 
integrals becomes very large necessitating huge RAM or disc storage. Different CPUs accessing the same memory or 
disc 
will be very time consuming. The best policy is to divide the calculation by regions and attribute each region to 
different 
nodes since each point can benefit from the wave function computed in an earlier calculation.

\subsubsection{Methods for computing {\sl ab initio} points}

\paragraph{Non reactive scattering (elastic and inelastic)}

Computing {\sl ab initio} points for non-reactive scattering has not been extensively discussed during the course 
of the 
colloquium/workshop, since it was felt that the situation is quite favorable. Being it for interaction with H, H
$_2$, 
or 
He, methods based on the CCSD(T) {\sl ab initio} suites \cite{Bartlett2007RevModPhys.79.291} are considered today 
to be 
of sufficient quality, 
especially so for non-asymptotic distances. The only problem remaining is the size of the atomic bases. 

Computing inelastic S-matrix elements (and even more so for elastic $S$-matrix elements, required for pressure 
broadening 
or momentum transfer cross sections), necessitates having a PES precise of the order of a few percent, and a 
well-calibrated long-distance behaviour. It has been shown repeatedly that the CCSD(T) method must be used with 
care in 
order to achieve this level of precision: proper correction of the Basis-Set Superposition Errors (BSSE, 
\cite{vanDuijneveldt1994doi:10.1021/cr00031a007,Mentel2014doi:10.1021/ct400990u}), large and flexible enough 
atomic basis sets are needed. An example is the so called 
correlation consistent series 
of Dunning \cite{Dunning1989JChPh..90.1007D} (aug)-cc-pVNZ, with N running from D (double-$\zeta$) to 7 ($7-\zeta
$) 
\cite{Feller2000JChPh.112.5604F}.  With such large sets of atomic basis sets, 
extrapolations towards infinite-basis sets are possible, even if quite heuristic (see e.g 
\cite{Okoshi2015JCC:JCC23896} 
and references 
cited therein). 

There exist new perturbational schemes that take explicitly into account the wave-function cusp for vanishing 
inter-electronic distances
between electrons. The newest versions
 (so-called F12 codes, \cite{Hattig2012doi:10.1021/cr200168z,Kong2012doi:10.1021/cr200204r}) are now reliable. To 
summarize, employing a properly balanced basis set, and taking into account electronic correlations via a CCSD(T)-
F12 
method seems a satisfactory way to compute the non-binding interaction of He/H/H$_2$ with the molecule under 
study. It 
is 
not yet fully agreed upon, whether this method is sufficient at all distances, or whether a method especially 
suited for 
long distances (like SAPT, \cite{Jeziorski1994doi:10.1021/cr00031a008,Misquitta2005JChPh.123u4103M}, or else an 
{\sl ab 
initio} computations of multipoles and polarizabilities, 
like in \cite{Bussery-Honvault2008JChPh.129w4302B,Bussery-Honvault2009JPCA..11314961B} ) is more economical and/or 
more 
precise.

Many examples of PES are now available, showing that the potential energy wells are well behaved and of modest 
depth 
(less than a few hundreds of cm$^{-1}$).  Exceptions are some
atom-H$_2$ interaction (for example the \ce{C(^3P) + H_2},  with the \ce{CH_2(^3B_1)} radical in the PES), or some
ions-molecule interactions, like $\mathrm{HCO^+ - H_2}$, with a well depth of $\sim 1487\,\mathrm{cm^{-1}}$, 
\cite{Masso2014JChPh.141r4301M}. 

\paragraph{Reactive scattering}

When used to study reaction dynamics the Potential Energy Surfaces must fulfill certain requisites. They must 
cover all 
the accessible configurational space, from reactants to products; describe all the minima and saddle points with 
chemical 
accuracy. At room temperature $ T \leq 1$ kcal/mol $\approx 1\,\mathrm{mE_h}$, but, for lower temperature studies, 
the precision 
should be higher; accurately describe the dissociation channels as well as the reactants and products; and 
preserve the 
permutational symmetry of identical atoms.
      
 Additional care must be taken when studying low temperature interstellar chemistry. Interstellar chemistry often 
involves open-shell radicals as reactants and/or products. This implies special care in the description of the 
dissociation channels as the open-shell radicals usually present degenerate states, that produce quasi-degenerate 
PESs 
difficult to compute accurately due to convergence problems in the algorithms used. In addition, the energy 
differences 
between these PESs is of the order of the spin-orbit coupling which must be taken into account. These quasi-
degenerate 
electronic states, often cross each other producing diabatic PESs and coupling terms that can be important to 
study the 
reaction dynamics. Often the reaction proceeds on the lower PES with small, submerse or non-existent barrier, 
being the 
reaction dynamics dependent of long-range part of the potential. This should have the correct R$^{-n}$ behaviour, 
but 
the 
van der Waals coefficients involving open-shell systems are difficult to compute using the present available {\sl 
ab 
initio} codes.   
    
\subsubsection{Methods for fitting {\sl ab initio} points}

Fitting the {\sl ab initio} points onto one single functional used to be one bottleneck between the computing of 
isolated 
{\sl ab initio} points and the subsequent dynamics. Indeed, full dynamics, as opposed to Transition State Theory 
or on-
the-fly computing of potential energy� relies on a global functional expressing the whole PES noted as ${\cal F} 
(R, 
\Omega, r)$, $R$, intermolecular distance, $\Omega$, angles that set the relative attitude of one molecule with 
respect to 
the other, $r$, the collective intramolecular coordinates . It is remarkable that this bottleneck has not been identified as such in the various interventions of 
the 
colloquium/workshop. 

For inelastic scattering, {\bf\tt MOLSCAT} has all necessary subroutines for using the PES for each types of 
scattering 
( rotor/atom; rotor/rotor)\footnote{Help and references in the {\bf\tt MOLSCAT} site http:// www.giss.nasa.gov/
tools/
molscat/doc/ for example}. For {\bf\tt Hibridon}, the different types of PES are entered in a more flexible way. Help 
and references to found in the Hibridon site \texttt{http://www2.chem.umd.edu/ groups/alexander/hibridon/hib43/
hibhelp.html.} 
Also, 
fitting codes exist that systematically fit the {\sl ab initio} points onto the right functional forms for {\bf\tt 
MOLSCAT}, minimizing the errors and the number of needed parameters 

Also, Szalewicz {\sl et al.} devised an economical way to fit for non-reactive scattering, based on site-site 
expansions. This functional form has been treated for water-water potentials , and for large molecule (\ce{HCOOCH_3})
colliding with helium \cite{Faure2011JChPh.135b4301F}. Fully testing this economical way of describing the PES for 
inelastic scattering is desirable. 

\paragraph{Fitting points for reactive scattering PES }

Another approach to build the PES is to smoothly join the {\sl ab initio} point with cubic splines. The accuracy 
of this 
approach depends on the density of the data, but is unable to reproduce the long-range behaviour of the 
approaching 
species.

Despite all the improvements and experience of the last years, building a PES useful for reaction dynamical 
studies from 
{\sl ab initio} data is still an art. No universal recipe is available for such desiderata. Different options can 
be 
used 
for problems like:
\begin{itemize}
\item Coordinates. Jacobi coordinates useful for scattering calculations are not suitable to describe bond 
breaking. 
Inter-atomic distances, hyperspherical coordinates, scaled coordinates, etc., are multiple choices available for 
fitting.
\item Accurate description of the reaction fragments. One approach is the many-body expansion, where the total 
inter-
atomic potential is defined as a sum of atomic, diatomic, triatomic, etc., terms. Each term should go to zero as 
one of 
its atoms departs from the others. This procedure warrants the correct dissociation of all the reaction channels 
but 
becomes infeasible for large systems.
\item Permutation of identical atoms. The PES should be invariant to the exchange of identical atoms. This can be 
accomplished taking care of the coefficients of the polynomials or using symmetric coordinates to represent the 
potential.
\item Treating different regions which have different energy behaviour. Chemical regions such as deep wells and 
transition states behave different from long-range interactions. The first ones depend on orbital superposition 
and 
decay 
exponentially with the inter-atomic distances while the second ones should approach zero with a $R^{-n}$ 
dependence. How 
to join these regions? Using an energy switch approach or damping the $R^{-n}$ at small $R$ values.
\item Curve crossing. Potential energy surfaces involving radicals frequently presents crossing of electronic 
states, 
Renner-Teller effects and  Jan-Teller cusps. The treatment of those systems, building the adiabatic and diabatic 
crossing 
terms or fitting the adiabatic potentials are different ways to solve the problem.
\item The functional form. This is a question where there are a lot of options. The use of Morse-like functions 
with 
varying parameters, the use of polynomials damped for large $R$ values, the use of localized polynomials, etc., 
are 
choices available for the fitting of the {\sl ab initio} data.  
\end{itemize}

\subsection{Dynamics}
 
Once the PES has been computed and the isolated points fitted onto a suitable functional form, the next step in 
our program is to perform dynamics on the PES. Several cases occur, that have been discussed in the course of the 
workshop. The case of elastic/inelastic scattering is usually the easiest case, even if some noticeable 
difficulties 
may arise (see~\ref{sec:inel}). Full dynamical solution of the reactive scattering problem 
remains extremely difficult, and is usually confined to either tri-atomic (possibly tetra-atomic) 
reaction, and/or to direct reactions, with sufficiently short reaction times in order for time-dependent solutions 
of the 
Schr\"{o}dinger equation to remain practical.

\subsubsection{Inelastic scattering and applications}\label{sec:inel}
\noindent
Discussion lead by \textsc{ A. van der Avoird, L. Wiesenfeld}\\

Quantum dynamics of elastic/inelastic scattering has been developed and perfected over the course of several 
decades, alloying today to treat large and sophisticated problems. It includes many elastic/inelastic scattering 
problems relevant for astrophysics: from hydrides of general formula XH$_n$ (closed or open shell) to main 
molecules of interest to astrophysics, all in collision with He, H, H$_2$, see 
\citet{Schroer2005A&A...432..369S,Dubernet2013A&A...553A..50D,Roueff2013ChRv..113.8906R}. It must be underlined thata lot of effort has been 
devoted also 
to electronic collisions, which were not discussed during the course of this workshop \cite{Tennyson201029}.

All those systems fare considered simple: Rigid target, sufficiently small number of rotational levels excited, 
low enough temperatures at which the rates $k(T)$ have to be computed. For those systems, it is considered that 
existing codes, either fully converged or with approximations that are known and under control, suffice for all 
practical purposes. It has been underlined that nowadays, elastic/inelastic scattering may be pushed towards very 
sophisticated problems, computed fully ab initio, as this workshop implies: heavy-heavy scattering (like CO-CO), 
pressure broadening, collision induced absorption, presence of external fields, very low $T$, down to the mK 
level.
Recent progresses include inclusion of bending/stretching for small systems (like HCN, CO), some internal motion 
excitation (like \ce{CH_3OH}).

However, the sophistication of present programs is such that the barriers towards more complex now are of 
computational nature. Those problems include, many more rotational levels (for heavier molecules, like 
\ce{HCOOCH_3}, HNCO, \ce{HC(O)NH_2}), full quantum computation of doubly inelastic sections (both projectiles and 
targets 
undergo transition), bending modes of molecules like H$_2$O, converging elastic cross sections towards higher 
temperatures, in order to compute transport properties or pressure broadening  \cite{Faure2013JQSRT.116...79F}\cite{frommhold:book}. Clearly 
efforts should be undertaken for being able to solve time independent Schr\"odinger equation with representative $N
\times N$ matrices with $N\gtrsim 5000$.

\subsubsection{Far from perfect: deep wells, bimolecular reactions}
 
\begin{figure*}[htbp]
\begin{center}
\includegraphics[width=0.7\textwidth]{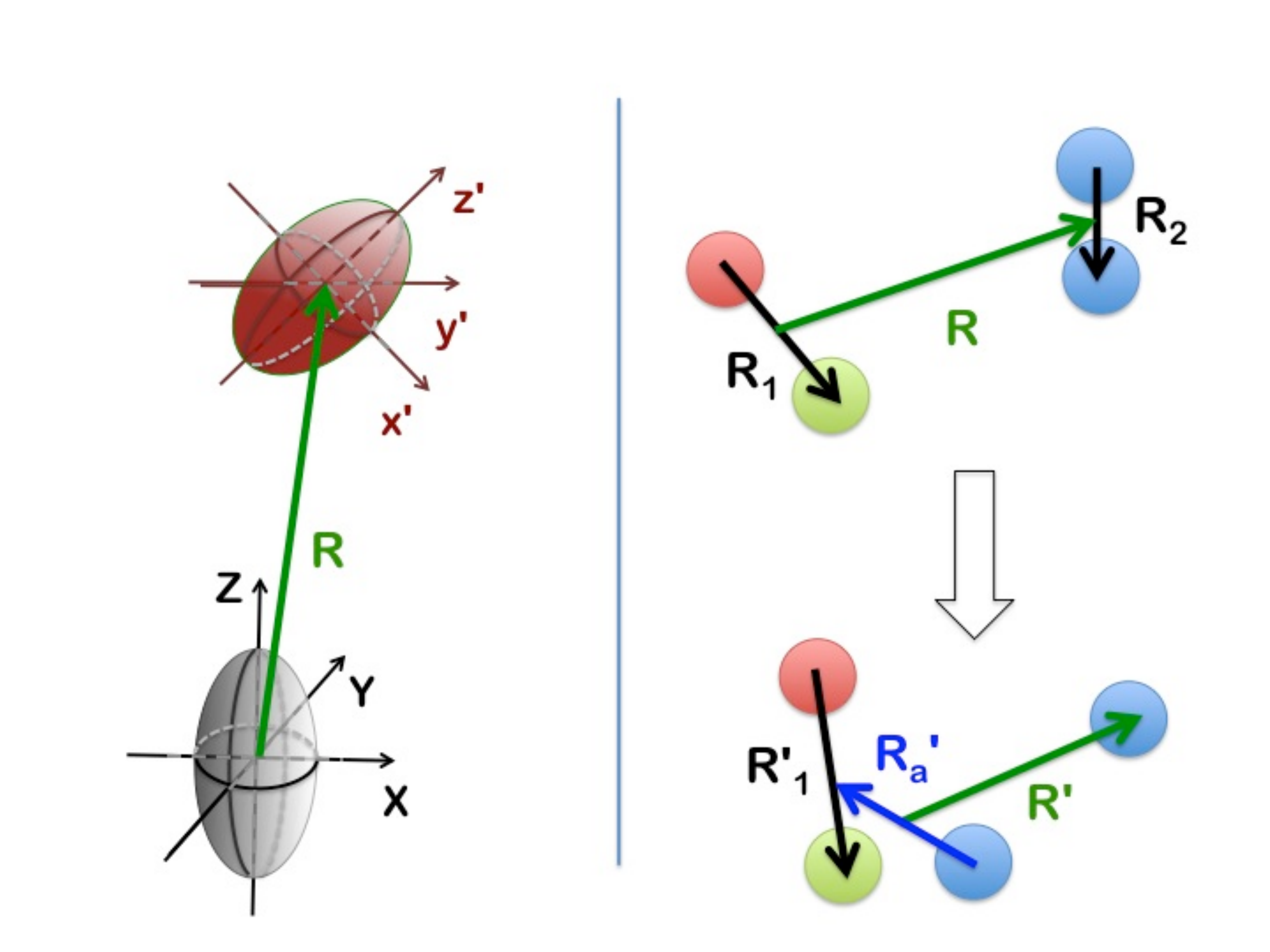}
\caption{Figure schematizing the difference between elastic/inelastic scattering (Left) and reactive scattering 
(right). 
On the left (scattering), the two ellipsoid retain their character during the scattering process. One set of 
coordinate 
e.g. $\mathbf{R}$ ($R,\,\theta,\, \phi$ in the OXYZ frame) as well as the orientation of the $Ox'y'z'$ (e.g;, 
three 
Euler 
angles $\alpha\,\beta,\,\gamma$. On the right hand side (reactivity), the asymptotic systems of coordinates change 
during 
the event, as exemplified by the figure.}
\label{fig:coord}
\end{center}
\end{figure*}

Computing of reaction states, branching ratios has been a major challenge  for many years, even for three atoms 
reacting  
in a simple way. Many complications arise, that make these computations difficult:
\begin{enumerate}
\item \textsl{Definition of the coordinates}\hspace{1.5em}
For inelastic scattering, one single system of coordinates suffices to describe the scattering for $-\infty < t < 
+
\infty $, see figure~\ref{fig:coord}, left panel. One single expansion of the potential, one single propagation at 
energy $E$ of the Schr\"odinger wave function is enough to get the $S(E)$-matrix, hence all imaginable 
observables. This is
not at all true for reactive scattering as is exemplified in right panel of figure~\ref{fig:coord}. One sees that 
at $-
\infty < t< t_0$, the coordinate system is adapted in a different way (coordinate $\mathbf R$ describes collision) 
then 
for $t_0 < t < +\infty$ , where a different $\mathbf R'$ is the scattering coordinate. It is possible to properly 
define 
so-called democratic coordinates that put all configurations on a same footing, for up to 4 centers, but at the 
expense 
of very cumbersome analytic wave function bases (see section \ref{sec:H5}) and difficult description of the 
asymptotic channels.

\item \textsl{Multiple surfaces}\hspace{1.5em}
While some reaction occur on one and single surface (some rearrangement reactions for singlet molecules, for 
example), in 
many cases of relevance in astrochemistry, ions and radical species are reactants. In those cases, a good 
description of 
the reaction event, including all channels and quantum numbers becomes a major task, very seldom put to fruition 
\cite{QUA:QUA24661}. Several problems appear at the same time:
\begin{enumerate}
\item{Spin effects.} When some of the species are open shell, all surfaces split according to the total electronic 
multiplicity. The degeneracies is usually different in the asymptotic channels and in the reactive part of the 
potential 
energy surfaces (PES), and the aspects of the various multiplets PES are vastly different. It means that  
branching 
ratios may be computed only by fully taking into account the spins. Furthermore, weak magnetic / spin-rotation 
terms 
allow for intersystem crossings, increasing further the complexities of the quantum treatment.
\item{Conical intersections and higher order singularities.} Spin or charge  result in several PES's describing the 
reaction. 
Generically (meaning, in a non-specific case, like identical particle effects) these surfaces intersect in several 
ways: 
avoided crossings, conical intersections, and for $n>3$ atoms,  higher order singularities, like intersection of 
conical 
lines. These lines/points make it difficult or impossible to propagate semi-classical trajectories, and make it 
necessary to 
resort to brutal approximations when propagating classical trajectories. It is very difficult to propagate 
wave-
packets and even time-independent wave functions in such topologies. There also the need to properly define diabatic vs. adiabatic 
PES's, a problem with no unique solution.
\end{enumerate}
\item \textsl{Indirect reactions}\hspace{1.5em}
For $n=3$ atoms, with a PES of one sheet,   codes exist, capable of computing the $S$-matrix, like the ABC 
code. Some recent extension even allow some indirect reaction to take place \cite{Werfelli:2015jk}, but at the expense of a great 
numerical 
effort. An interesting list is published by the CCCP6 group (http://www.ccp6.ac.uk/downloads.htm), even if 
partial.
\item \textsl{Tunneling}\hspace{1.5em} At low temperature, a few K, reactions may progress by tunneling, 
especially so 
if H or possibly D atoms are involved. This is a well-known characteristic of astrochemistry, and very low-$T$ 
chemistry 
in general. Inclusion of tunneling has been described by various formalisms, the instanton  being one of them \cite{Zhang:2014qv}. 
Including those effects that go way beyond classical Transition State Theory is one of the goals of quantum TST, 
see 
below.
\end{enumerate}

The situation seems thus very complex, and it appeared that we are far from being able to tackle the quantum 
reactivity in its full extent. We have neither a full algebraic/analytical picture, nor a set of algorithms capable 
of 
treating actual, interesting cases. The MCTDH (Multi-Configuration Time Dependent Hartree) formalism 
\cite{Meyer:2009ly,QUA:QUA24661}  is 
very 
powerful for single surfaces --even multidimensional-- and not too long interaction times. The ABC program does 
the same for three 
atoms in a time-independent case. Some formalisms keep on appearing, but except for very large brute force 
computations, 
it does not seem, in the editor's opinion, that a breakthrough has recently been reached, for a full quantum 
description of reactive scattering, even if some may argue that the so-called ``polymer-ring'' formalism is due to play a very interesting role \cite{Craig2005JChPh.122h4106C,Suleimanov2013CoPhC.184..833S}.

\subsection{ Transition states}

\begin{figure}[htbp]
\begin{center}
\includegraphics[width=0.43\textwidth]{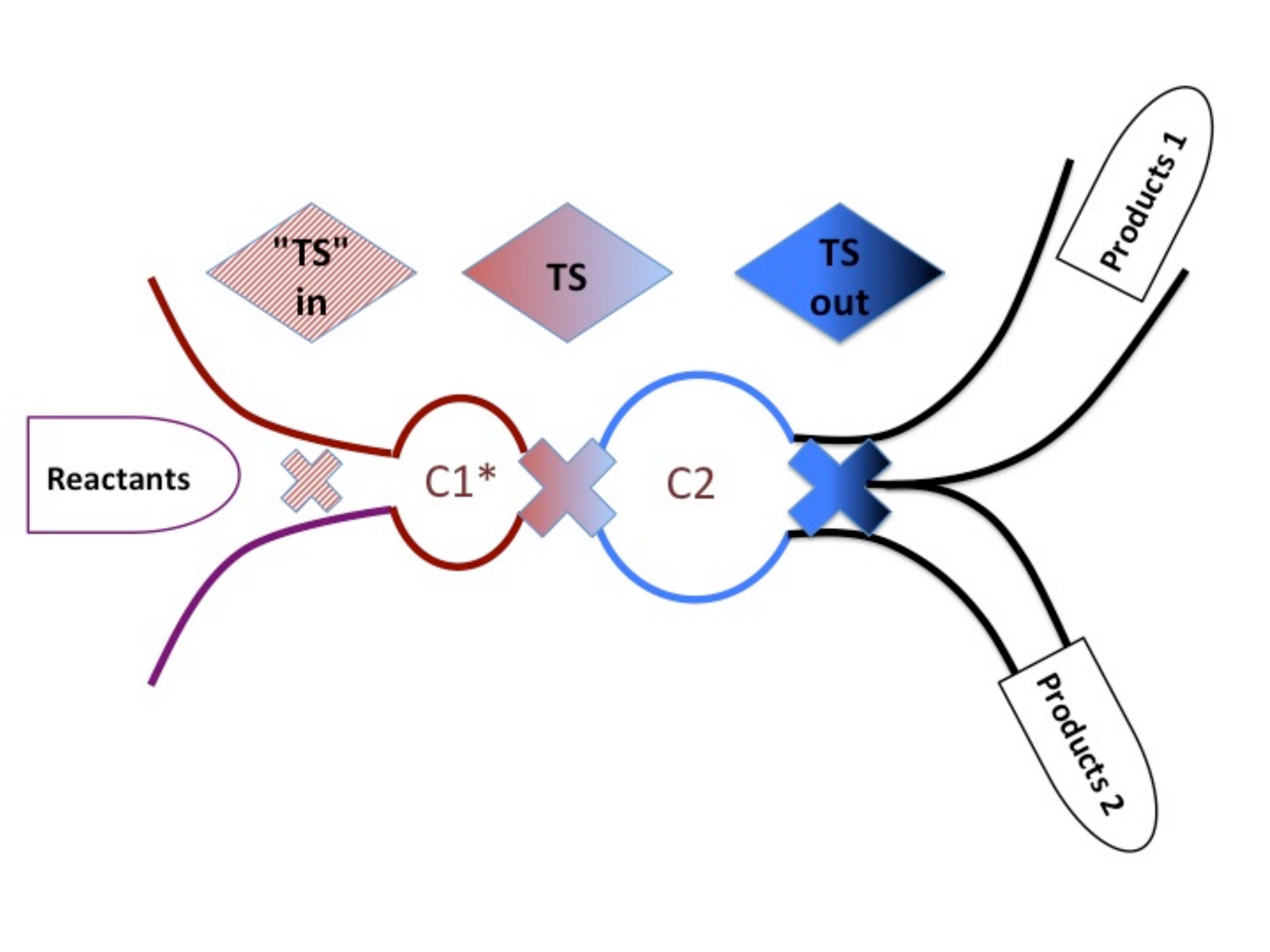}
\includegraphics[width=0.43\textwidth]{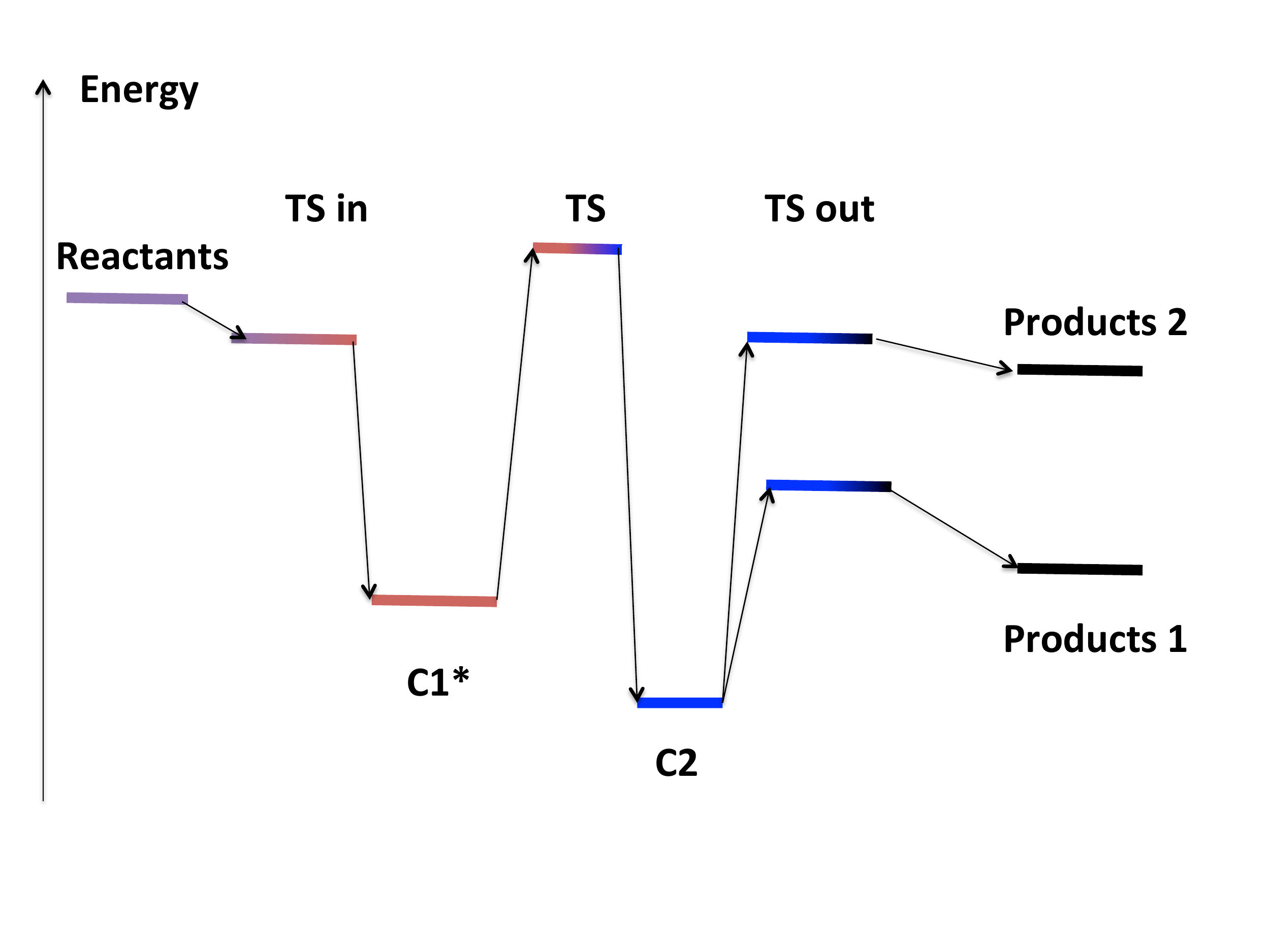}
\caption{A scheme of the progression of a reaction similar to figure-\ref{fig:scheme1_1}, from reactants, to two successive sets of 
complexes, an excited one  $\mathrm C_1^*$, a metastable one C$_2$, and various exit routes. This type of scheme 
is very 
usual, and more on the simplified side here. Upper panel: A schematic view in some abstract phase space. Each step 
in 
this simplified model is separated from the next by a transition state -depicted by crosses. Lower panel: More 
usual 
picture with some reaction coordinate as horizontal axis and energy as a vertical axis. Same nomenclature for both 
panels. In some cases, several parallel routes exist.}
\label{fig:TST-1}
\end{center}
\end{figure}

Repeatedly in the course of the workshop,  the concept of transition state 
 theory (TST) appeared as one of the main tool able to be substituted to exact dynamics when it is obvious 
that numerics becomes impractical or unfeasible. TST is able to (i) conceptualise the outcome of a chemical 
reaction (ii) quantify 
the chemical rates, even if in an approximate way. While transition state theory   is by no means a novel idea 
(it dates from 1913, see \citet{Fernandez-Ramos:2006aa}), its mathematical and chemical aspects have greatly been 
improved and extended. Both these aspects were of interest in the context of this workshop:  The progress made in 
the mathematical community and the actual use in non trivial contexts, like radical-radical reactions.
  
   The usual way to define a TS is to determine the minima of flux of probability density, from the reactant 
region to 
the product region. These minima of flux are considered as bottlenecks, and serve as supports for the rates to be 
introduced in the 
master equations for the overall chemical networks. 
Reactant and product regions are defined in a different way for mathematical aspects (regions of representative 
phase 
space) or for a quantum approximate view (quantum states of the reactants/products), figure\ref{fig:TST-1}. Note 
that a semi-classical TST is also part 
of the present program, based either on some Wigner picture of phase space, or, more simply,  taking into account 
zero-point energies and elementary tunneling effects, possibly via instanton type of approximation.
 
There are several important features of TST, that make it the most attractive way to compute rate of chemical 
reactions, in many cases. In all common applications, the TST is based not on a full PES determination, long and 
painful, but rather simply on finding the stationary points of the PES. Many ab initio codes are very well suited 
for finding semi-automatically stationary points $P_i$. One proceeds then as follows, in a way compatible with a 
fully automated feature of the quantum chemistry code: (i) Form the Hessian $\mathsf H(P_i)$ of the potential at 
the stationary point (symmetric matrix of second derivatives with respect to coordinates (We work here 
only in configuration space); (ii) Find its pairs of frequency eigenvalues, and concentrate on those $P_i$ where 
all 
eigenvalues pairs are real, but for one which is imaginary; (iii) Let the imaginary pair be associated with 
eigenvectors $\xi_i$; this $\xi_i$ is the local reaction coordinate; (iv) all other coordinates are bath 
coordinates and serve to define the density of states at the TS; (v) this density of states, either at fixed 
energy or temperature, defines the flux from reactants to products. All details may be found in many references 
and textbooks (Mathematical context: \cite{Wiggins:2003uq}; physical chemistry context: 
\cite{Klippenstein:2014xy}). This program is perfectly adapted for many reaction schemes, and has been used for decades.

 Also, TST, being such an intuitive concept, and being defined both at fixed energy (micro canonical ensembles) 
and at fixed temperature (canonical ensembles), its domain of definition has been extended to many abstract 
dynamical systems.
 
 As a final point, it is obvious that TST is perfectly adapted for obtaining approximate (even with a very good 
approximation, see
section \ref{sec:TSTprac}) reaction rate in the case where: the whole reaction is exoergic --$ E(\text{products}) 
< E(\text{reactants}) $-- and where the energy of the TS is such that $E(P_i)> E(\text{reactants})$. Many of the 
rates proposed in databases for combustion \cite{Baulch:2005aa} or astrochemistry (KIDA,
 \texttt{http://kida.obs.u-bordeaux1.fr/}) are based on TST at high temperature and finite pressure, even if these conditions are not fully 
relevant to 
many 
astrochemical environments. Many examples are vividly illustrated in chemical schemes showing the relative 
energies various asymptotic channels, reactants and products, TS and stable or meta-stable intermediates 
\cite{Kaiser:2002fk}\cite{Li:2013uq}.

Astrochemistry relies on this definition of TS for many reaction schemes, even if it is not at all fully 
applicable. One of the way to salvage TST (or actually, it was one of the ways that led to its discovery) is to 
add the centrifugal barrier  to the PES. For ion-neutral interactions, this leads to Langevin-like 
approximations, where the capture cross sections and rates depend on the multipole/polarizability of the neutral 
species. For neutral-neutral attractive PES (no barrier or low barrier at entrance), it leads to other 
formulations of capture theories, like \citet{Faure2000CP....254...49F,Georgievskii:2005aa}.

Clearly, astrochemistry, lacking thousands of reaction rates at temperature ranges from 10 to 1000~K, is in dire 
need of viable TST, which would be valid even if \emph{no entrance barrier is present}, like for most 
ion-molecule, many radical-radical reactions and some radical-neutral reactions. 

Note that in its usual forms TST predicts total reaction rates, while it often desirable to get branching ratios 
between the various products. Let us think of a very common model like: $\mathrm{ A + BC\leftrightarrow ABC^* 
\leftrightarrow AB+C \, / \, AC + B}$. The stable compound product ABC is produced from the excited form ABC$^*$
 by third body collision (pressure effects) or else photon emission, usually a slower process than 
decomposition. Hence, knowing the branching ratios is by no means a trivial task, experimentally or 
theoretically. 

\subsubsection{Mathematical grounding} \label{sec:TSTmath}
 
\par\noindent 
Text by  \textsc{T. Komatsuzaki, H. Teramoto, M. Toda, and L. Wiesenfeld.}\\

For most of the discussion, TST in mathematics is defined within the framework of \emph{Hamiltonian classical 
dynamics}. 
While for $n=2$ degree of freedom (dof) systems (such as triatomic linear reaction),  rigorous (classical) TST 
exists, for $n>2$ degrees of freedom, the formalism is still under development. The energy domain at threshold or 
just 
above is now clearly understood (see below), but as soon as energy increases, all possible scenarios for the 
dynamics in 
the vicinity of the TS are not fully known, nor fully described. Let us state those points in more details.

A definition of TS at threshold for 2 dof was  put forward 
some decades ago, based on periodic orbits \cite{POLLAK:1978aa} . For $n>2$ the existence of suitable dividing 
manifolds  in the Hamiltonian flows, while known for some years too \citep{FENICHEL:1971aa,MACKAY:1990aa}, has 
been 
fully 
appreciated only more 
recently \citep{Komatsuzaki:2001aa,Wiggins2001PhRvL..86.5478W,Waalkens:2004aa}. The development of so-called 
"Normally 
Hyperbolic Invariant Manifolds" (NHIM), its application to capture or unimolecular reactions steadily improved the 
understanding of Hamiltonian dynamics and its relation with TST.

The main property that classical dynamics fully clarified is the notion of 'no-return' TS. The NHIM's are defined 
as 
manifolds (hyper-surfaces) that act as ``generalized saddle'' in the phase space at which the system perpetually stays even with finite momenta and nonlinear couplings among the modes
at threshold. The NHIM being an unstable equilibrium set, it has stable 
and 
unstable manifolds\red{. The} codimension \red{of the stable and unstable manifolds} is 1 \red{at} the energy\red{. The property of codimension one is of crucial importance to define the state or domain of reactants and that of products. Codimension $n$ means a dimension less than $n$ from the dimension of the ambient space. For example, if the dimension of the ambient space is three, in order to divide the space one must have a space of two-dimension, i.e., a plane or surface. Likewise, if the dimension of the space is two, one must have a space of one-dimension, i.e., line. Otherwise, one cannot divide the space into two. It is known that the codimension one stable and unstable manifold emanating from the NHIM serve as the so-called reaction tube or conduit so that  }
at \red{the} energy $E$,  trajectories representing the shape of the molecules and originating on the reactant side \red{move} all the way \red{necessarily through the reaction tube} to the 
products side\red{.} Hence, measuring the flux of trajectories across \red{a surface defined to intersect vertically the reaction tube in the phase space, i.e., serving as (rigorous) no-return TS} 
amounts to measure the 
reaction flux at the energy $E$. 

The dynamics at the threshold of the TS is fairly well characterised by now, including the integrability of the 
motion and various ways of introducing angular momentum \cite{Kawai:2011aa},
 \cite{Wiesenfeld2003JPhB...36.1319W},\cite{Ciftci:2012aa}.
 
Three  avenues of progress are pursued now (let $E^*$ be the threshold energy):
\begin{enumerate}
\item What is the fate of \red{NHIM (that provides a natural definition of a no-return TS)} 
at energies $E>E^*$? How does it survive, how does it bifurcate? Can 
we 
learn something from the well-known 2dof case? 
\item Building on the preceding item, is there a way to go from a micro-canonical picture (flux et fixed $E$) to a 
canonical picture, 
yielding classical rates $k(T)$ directly, like it is doable numerically?
\item How is it possible to go from some kind of unimolecular/capture type of description towards a full 
description of 
a 
reaction, with multiple asymptotic channels and TS's. 

\end{enumerate}

\begin{figure}[htbp]
\begin{center}
\includegraphics[width=0.45\textwidth]{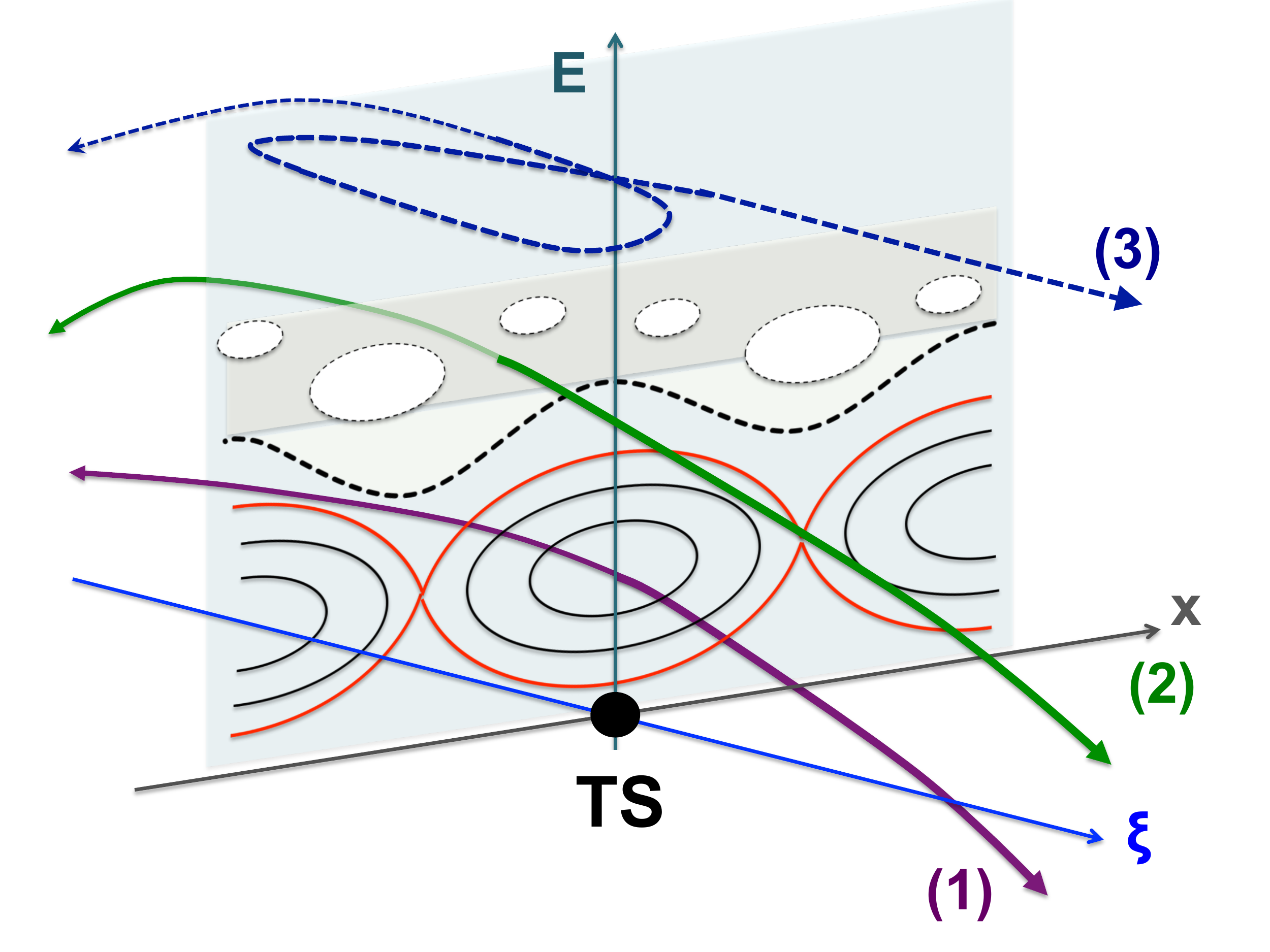}
\caption{An abstract view of \red{TS},
beginning at threshold
and extending up in total 
energy $E$. 
If we have $n$ degrees of freedom, the $\xi$ coordinate in the figure labels  $\xi, p_{\xi}$, the canonical 
coordinates of the local reaction path. $x$ labels the set $\left(x_i,\,p_{xi}\right)$, $i=1,n-1$, the bath 
coordinates. 
The trajectory (1) crosses the 
\red{TS}, 
when motion locally while crossing
is fully regular, \red{corresponding to the existence of approximate invariant of action variables, i.e., good quantum numbers).}
Trajectory (2) higher in $E$ \red{starts to experience chaos because of nonlinear couplings among the modes but yet free from 
recrossing because resonance is impossible between bath coordinates and reaction coordinate $\xi$ whose frequencies are real and imaginary, respectively. }
For trajectory (3) \red{much higher in $E$ where chaos is developed so that one cannot deal with nonlinear couplings perturbatively. 
Reaction dynamics at such high energies are one of the central subjects in nonlinear dynamics theory of reactions. See for example, Ref.
\onlinecite{Li2006,Teramoto2011, Teramoto2015, Teramoto2015b}}}.
\label{fig:scheme-tst}
\end{center}
\end{figure}

The discussion in the workshop made it clear that, while the Hamiltonian dynamics pictures begin to emerge in a 
coherent 
way, applications towards actual reactions are still lacking, probably because of the  widely different points of 
views, languages and knowledge bases of mathematical or molecular dynamicists.

\subsubsection{Quantum Transition State Theory}
\par\noindent
Text by \textsc{H. Waalkens}\\

Especially for reactions involving light atoms quantum effects might become important. Even with the computer 
power that 
we have today full-fledged ab initio quantum computations are often not feasible. In recent years it has been 
shown that 
the geometry underlying classical reaction dynamics and the algorithms to compute the geometric structures that 
govern 
reactions can be utilised and extended in such a way that quantum reaction rates can be computed very efficiently.  
The 
classical phase space structures which govern the reaction dynamics through a phase space bottleneck induced by a 
saddle 
equilibrium point can be computed in an algorithmic fashion from a Poincare-Birkhoff normal form expansion. The 
Poincare-
Birkhoff normal form leads to a canonical transformation to new phase space coordinates in terms of which the 
classical 
dynamics locally decouples into a saddle (i.e. reaction) degree of freedom and centre (i.e. bath) degrees of 
freedom.  
The new normal form coordinates give explicit expressions for the various phase space structures which control the 
classical transport across a saddle. This includes a recrossing free dividing surface and the directional flux 
through 
this surface is . The dividing surface is spanned by a normally hyperbolic invariant manifold (NHIM). The NHIM has 
stable 
and unstable manifolds which extend away from the saddle point into the reactants and products part of the phase 
space 
and channel reactive trajectories from reactants to products (and vice versa). The knowledge of which regions they 
sweep 
out in the reactants and products region is the key to understand state specific reactivities. The normal form 
expansion 
is of local validity and hence allows one to compute  at first only the local pieces of the stable and unstable 
manifolds 
near the saddle. The local pieces can then however be grown into the reactants and products region by letting the 
flow 
of 
the original (untruncated) dynamical equations act on them.

In the semiclassical limit the classical phase space structures that govern reaction dynamics form the backbone 
also for 
the quantum mechanics of reactions. Similar to the classical case the quantum dynamics can also be locally 
decoupled to 
any desired order into saddle (reaction) and centre (bath) degrees of freedom.  This can be achieved in terms of a 
quantum normal form expansion which yields a unitary transformation which 'locally' simplifies the Hamilton 
operator in 
the neighbourhood of the saddle to any desired order. The expansion can be cast into an algorithm using the 
Wigner-Weyl 
symbol calculus. The main difference to the classical case is then that the Poisson bracket being replaced by the 
Moyal 
bracket.  What  'local' simplification means can be understood from the theory of micro local analysis. The 
quantum 
normal form allows one to compute quantum reaction rates with high precision (it takes full account of, e.g., 
tunnelling 
effects). It also allows one to compute very efficiently the complex energies of Gamov-Siegert resonance states 
which 
describe the decay of quantum wave packets initialised on the classical NHIM. 

The quantum normal form can be considered to be a rigorous realisation of quantum transition state theory. 
Although the 
efficiency of the quantum normal form has been demonstrated for various systems that are also still many 
challenges and 
open question: 
\begin{itemize}
\item Similar to the classical case the quantum normal form leads to an asymptotic series which has to be 
truncated at a 
suitable order. It is only valid in the neighbourhood of the saddle. In the absence of resonances between the 
centre 
degrees of freedom the normal form leads to integrable dynamics. Similarly the quantum dynamics resulting from the 
(truncated) quantum normal form results in as many commuting observables as classical degrees of freedom (i.e. all 
the 
centre degrees of freedoms have good quantum numbers). For high energies, it might no longer be reasonable to 
approximate 
the dynamics on the NHIM in terms of an integrable dynamics. This is similarly true in the quantum case.  
Especially it 
might happen that the NHIM bifurcates or even get destroyed.  Recently the case of  Morse bifurcations of NHIMs 
has been 
studied classically. It highly desirable to also understand this in the quantum case. 

\item Recently it has been shown how the classical normal form expansion can be achieved in the presence of rotation-
vibration coupling. The challenge here was that the symmetry reduced phase space of the N-body system given by the 
atoms 
that form a molecule is given as a quotient space for which it was unclear how to define canonical coordinates 
which 
form 
the starting point for  a classical normal form expansion.  It has been shown that such canonical coordinates can 
indeed 
be constructed and that the classical normal form expansion can then be carried out in the usual manner near a 
saddle. 
The difference in time scale between rotations and vibrations suggest however, that it is better to consider the 
saddles 
arising from rotation-vibration couplings from a nonlocal perspective. In fact the classical phase space 
structures that 
occur in this case should be considered again as Morse bifurcation of NHIMs. It is still an open question on how 
to do a 
quantum normal form for the case of rotation vibration coupling as such and in particular also taking into account 
the 
Morse bifurcations as already mentioned above.

\item Another question that has not been tackled yet for the case of quantum reactions using the semiclassical 
approach 
mentioned above is how to deal with reactions across a succession of two or more saddle points. In 
\cite{Waalkens:2008kx} it has been argued that this situation can be approached using the technique of quantum 
Poincare maps introduced in \cite{Bogomolny:1992fk}.  This however has not been pursued yet. 

\item For the classical case,  it has been shown how the coupling of a reacting system to a heat bath (described 
in 
terms 
a Langevin dynamics framework) can be taken into account for the classical normal form. For the quantum case, it 
is 
still 
an open question how to incorporate the coupling to a heat bath. 

\item Another problem that has intensely been discussed in the context of reaction dynamics in recent years is the 
occurrence of phase space bottlenecks not induced by saddles (in some context such bottlenecks are referred to as 
entropic barriers). Such bottlenecks appear to be in particular relevant for roaming dynamics. For two degrees of 
freedom, there are NHIMs formed by unstable periodic orbits associated with such bottlenecks. It is still an open 
problem 
how to treat such bottlenecks quantum mechanically.

\item Finally it is worth mentioning that one important prerequisite for the quantum normal form machinery to be 
applicable is the validity of the Born-Oppenheimer approximation. It would be highly desirable to also develop a 
quantum 
normal form to take into account non-adiabatic effects due to a conical intersection.  
\end{itemize}

\subsubsection{Practical implementation} \label{sec:TSTprac}

Text by \textsc{S. Klippenstein.}\\

For astrochemistry, much of the focus is on barrierless reactions since those with a significant barrier are too 
slow to 
be relevant at low temperatures. For barrierless reactions there is no saddle point on the PES (or if there is one 
it is 
submerged below the reactant energy), but a TS still exists as a dynamical bottleneck in the reactive flux. The 
location 
of the TS then arises from a balance between competing variations in the entropy and the enthalpy. This location 
varies 
dramatically with temperature, or equivalently, energy and angular momentum. 

The variable reaction coordinate (VRC)-TST approach treats the intermolecular dynamics separately from the 
intramolecular 
dynamics, with the latter assumed to be vibrationally adiabatic. Meanwhile, the intermolecular dynamics is treated 
via 
variational TST employing phase space integration to accurately evaluate the TS partition functions for variably 
defined 
reaction coordinates. For a given PES, the VRC-TST method reproduces trajectory evaluations of the flux to within 
about 
10\%.  

In general there are two distinct TS regimes. A long-range region, with a TS determined by long-range terms in the 
PES, 
determines the rate at low temperature. At higher temperatures, a short-range entropically driven bottleneck 
becomes 
important. Such a two TS picture is clearly seen for radical-molecule reactions,  but also arises in many radical-
radical 
and ion-molecule reactions.  

The accurate prediction of rates with the VRC-TST method relies on the availability of an accurate description of 
the 
intermolecular PES. For radical-radical reactions, single reference methods are not applicable, and multi-
reference 
methods such as CASPT2 or MRCI are generally used.  Direct sampling of ab initio energies, with on the order of 
10,000 
evaluations required for reasonable convergence, provides an effective means for evaluating the requisite phase 
space 
integrals. For ion-molecule reactions, long-range expansions are appropriate for a fairly broad range of 
temperature.
However, at high enough temperature, short-range effects become important and direct ab initio sampling is again 
effective. 

For many reactions, the long-range portion of the potential energy surface is fairly complex, with multiple 
orientational 
minima correlating with a variety of final products. This complexity leads to what has come to be called roaming 
dynamics. The separation into conserved intramolecular and transitional intermolecular modes, which is at the 
heart of 
the VRC-TST approach, also provides an effective approach for treating the dynamics and kinetics of roaming 
reactions 
\cite{Fernandez-Ramos:2006aa}\cite{Sivaramakrishnan:2010aa,Harding:2012aa}.

\subsection{The specifics of H$_n^+$ and analogous cases}\label{sec:H5}

Text by \textsc{O. Roncero}\\

	Reactive collision rates are needed to properly account for the observed lines in different astrophysical 
environments and determine the physical conditions. Hydrogen is the most abundant element in space and low 
temperatures 
are typically found in molecular clouds. In this environments quantum effects are typically important. For this 
purpose 
the formation rates of simple hydrides like CH$^+$ and OH$^+$, some of which were recently observed after the 
launch of 
the Herschel Space Telescope in 2010, can be   obtained with quantum methods 
\cite{Zanchet2013ApJ...766...80Z,Gomez2014ApJ...794...33G}. Some of these molecules are very reactive and do not 
have time to thermalize, and 
therefore  the state-to-state reactive rates determine the emission intensity of excited 
rotational states. 

For reactive collisions the two main "exact" quantum methods nowadays available  are the time-dependent wave 
packet 
method and the time-independent close coupling method. These last methods require the resolution of a set of 
coupled 
equations depending on a scattering coordinate. While feasible in inelastic collisions, provided that the number 
of 
coupled equation is not too large, for reactive scattering there is the additional problem of coordinates. For 3 
atoms, 
the hyperspherical coordinates are well adapted and there are several codes nowadays available for reactive 
scattering. For larger systems the hyperspherical coordinates, as those defined for 4 
\cite{Kuppermann1997doi:10.1021/jp9708207} and 5 atoms \cite{Kuppermann2011C0CP02907F} are 
difficult to implement, specially because the singularities of the Hamiltonian in those coordinates require the 
use of 
exact or approximate hyperspherical harmonics, which are difficult to obtain. In addition, the number of coupled 
channels increases enormously, making impractical their use even for 4 atoms systems.

Wave packet techniques seem to  scale better for larger systems. For few 4 atom systems the groups
of D. H. Zhang and more recently H. Guo \cite{Zhao2016JChPh.144f4104Z} are now able to get state to state cross 
sections 
but typically considering 
only few reactive channels. Also, for larger systems they have been applied using reduced dimensionality models. 
For 
rather large systems the MCTDH approach \cite{Beck2000PhR...324....1B} is a good alternative but the direct 
dynamics 
with this method is not well 
suited when long lived resonances are present. An alternative to use this method has been successfully applied by 
the 
group of U. Manthe for direct reactions with barrier  using the full dimensional MCTDH propagations and the 
quantum 
transition-state concept \cite{Welsch2015doi:10.1021/jz502525p}. However, for reactions involving long lived 
resonances 
and/or low temperatures (as 10K) all 
these methods fail.

The natural alternative is the use of classical methods. However,  quantum effects are important at low 
temperatures and 
in the presence of light atoms like hydrogen. Some "quantum" corrections need to be included, specifically to 
account 
for the zero-point energy and tunneling effects. For this there are a great number of semiclassical methods. A 
possible 
classification of some of these methods is as 'active', manipulating individual trajectories to overcome the 
problems, 
or 
"passive", in which trajectories non satisfying some criteria are neglected. 
Direct reactions are probably the best suited for being studied with classical methods. $\mathrm{H_2 + H_2^+ 
\rightarrow 
H_3^+ + H }$ reaction is one 
of such reactions, highly exothermic, by approximately 1.2 eV. In this case  two quantum effects, the Zero Point 
energy 
(ZPE) of the reagents and non adiabatic transitions, using the molecular dynamics with quantum transitions (MDQT) 
of 
Tully \cite{Tully1990JChPh..93.1061T}, were included in recent calculations using an accurate full dimensional PES 
with 
good descriptions of long 
range interactions \cite{Sanz-Sanz2015:/content/aip/journal/jcp/143/23/10.1063/1.4937138}. The results obtained 
are in 
good agreement with experimental results \cite{Glenewinkel-Meyer1997IJCH:IJCH199700039}, providing a good 
overall description of the process.
However, this is not always the case because individual trajectories can not reproduce interference, and for that 
some 
ensembles of classical trajectories need to be account for, as done in many semiclassical methods. These methods 
are 
typically difficult to be applied to multidimensional problems.
Recently, a whole collection of methods based on the path integral formalism have appear, as the ring polymer 
molecular 
dynamics \cite{Craig2005JChPh.122h4106C}, centroid dynamics \cite{Voth1989doi:10.1021/j100356a025}, etc, and are 
receiving an increasing attention in the community, as observed by the 
fast grow in the number of papers.  

At low temperatures and for systems presenting wells in the entrance channel, long lived complexes are formed. 
Under 
these conditions energy is redistributed among all degrees of freedom. The subsequent fragmentation can thus be 
described 
in many cases by purely statistical methods. The problem is to determine when direct mechanisms start having an 
important 
contribution, in detriment of the long-lived complex or statistical mechanisms. One example  is  
$\mathrm{H_2+H_3^+ \rightarrow H_3^+ + H_2}$ 
exchange reaction, for which experimental results \cite{Cordonnier2000JChPh.113.3181C,Crabtree2011JChPh.134s4311C} 
indicates that at low temperature (below 50~K) the results are 
consistent with statistical methods, accounting for nuclear spin permutation symmetry rules 
\cite{Hugo2009JChPh.130p4302H}. 

However, at higher 
temperatures (above 100 K) the experimental data indicates that the process is no longer statistical but direct. 
To 
describe the transition between these two mechanisms, a dynamically biased statistical method has been proposed, 
in 
which the statistical weight for the different rearrangement channels of products have been substituted by 
reaction 
probabilities  obtained with a quasi-classical method~\cite{ISI:000243891100017}. In this system the ZPE is larger 
than 
the binding energy, and it 
was found that when the full ZPE of H$_2$ and H$_3^+$ reagents are considered,  the ZPE is transferred to 
dissociating 
modes 
producing artificially too short lived complexes, the well known ZPE leakage, which gives non correct 
probabilities. To 
correct this ZPE problem, and considering that  the ZPE of the H$_5^+$ complex is similar  to that of the 
reagents, 
within 
10-20 \%, the ZPE of the reagents is reduced in order to avoid its artificial flow. To do so, using RRKM theory, 
it is 
found that to get similar quantum and classical density of states at the \ce{H_5^+} well it is needed to reduce the ZPE 
of 
reagents to only 25 \%. Doing so, the reaction probabilities did show a correct behavior, allowing to describe the 
transition between pure statistical mechanism at low temperatures and direct hop mechanism at high temperatures. 
Also, 
this simple model allows to determine the principal role played by ZPE in this kind of reactions. To get  more 
precise 
results, the ZPE effect has to be accounted for, and for these purpose RPMD calculations 
\cite{Suleimanov2013CoPhC.184..833S}
 are  performed now-a-days by Y. Sulemainov on this system.

\subsubsection{Quantum/classical and quasi-classical dynamics}

\noindent
Text by \textsc{A. Faure}

As explained above, semi-classical (or quantum-classical) and quasi-classical approximations provide interesting 
alternatives to quantum methods when the number of coupled-channels is exceedingly large. Semi-classical methods 
consists in solving the time dependent Schr\"odinger equation while in quasi-classical approximations the 
classical Hamilton equations of motion are integrated. Their common feature is that at least one degree of freedom 
(in general the relative motion) in the collision is treated classically. 

The idea of a mixed quantum/classical 
theory is rather old (1950s) but difficulties in the application of semi-classical theories (e.g. classical $S$-
matrix theory or WKB methods) have precluded them being routine computational tools. There is however a recent 
revival of interest for these theories and their application to inelastic scattering, in particular the mixed 
quantum/classical theory (MQCT) \cite{Babikov:2016fk}. On the other hand, quasi-
classical methods have been widely employed in reactive scattering and they have been also successful in inelastic 
scattering studies (e.g. \citet{Faure01082016}). In the quasi-classical trajectory (QCT) approach, 
batches of trajectories are sampled with random (Monte Carlo) initial conditions are they are analyzed 
statistically. State-to-state observables (cross sections or rate coefficients) are extracted by use of the 
correspondence principle combined with the bin histogram method. We note that more elaborate binning approaches 
exist, in particular the Gaussian weighting method developed by Bonnet and Rayez \cite{Bonnet1997183}. A 
particular advantage of the QCT method is that the computational time decreases with increasing collision energy, 
in contrast to quantum close-coupling methods. QCT calculations however ignores interference and tunneling 
effects. In addition, the correspondence principle cannot be generalized to the case of asymmetric top molecules 
owing to assignment ambiguities \cite{Faure:2004xw}. This particular drawback makes the 
MQCT approach of Babikov \& Semenov a very attractive and promising tool.

\subsection{Spectroscopy: A short view}

\noindent
Text by \textsc{L. Bizzocchi and C. Puzzarini}.

High-resolution molecular spectroscopy is a powerful tool to investigate various molecular properties that are 
critical 
in the context of astrochemistry. To fully exploit the potentialities of rotational spectroscopy in this field, it 
is 
necessary to know accurately the spectroscopic parameters of the molecules of interest. These parameters are: the 
transition (rest) frequencies, their intensities, the corresponding pressure-broadening and shift coefficients and 
their 
temperature dependence. For the relevant species, the parameters obtained by laboratory studies are then collected 
in 
databases that are continuously updated and improved.

The first and more important application of laboratory measurements is thus the retrieval of accurate rotational 
rest 
frequencies, which opens the way to the detection of new molecules in space. 

\begin{figure*}[htbp]
\begin{center}
\includegraphics[width=0.85\textwidth]{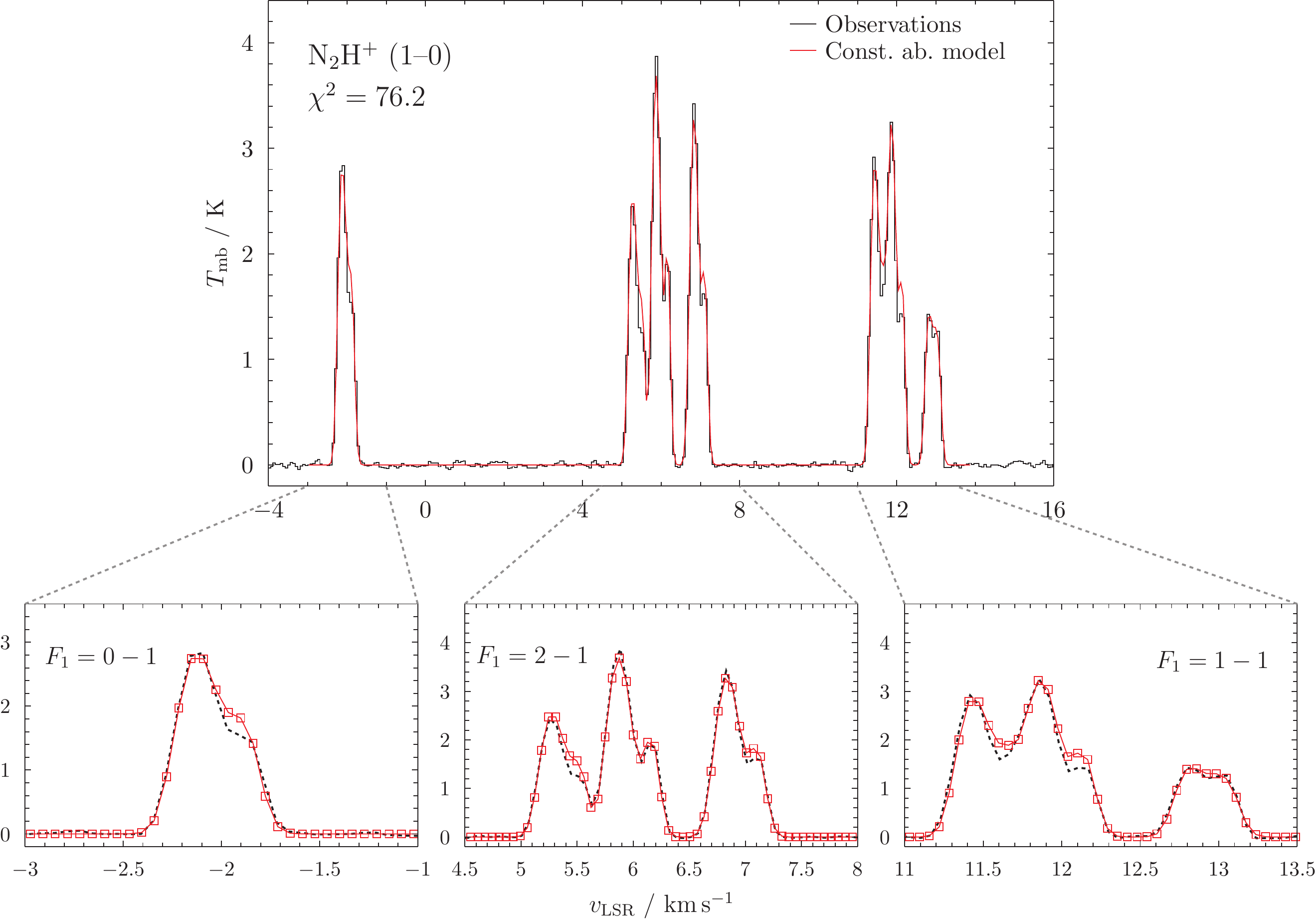}
\caption{Non-LTE radiative transfer modelling of the N2H+, J=1-0 transition
observed towards the pre-stellar core L1544.
The low turbulence of this cold and quiescent object allows accurate
determination of the radial velocity infall profile.
The histogram shows the observation data taken with the IRAM 30m telescope with 20 kHz frequency resolution.
The red curve shows the model spectrum computed using a refined cold
cloud model and constant molecular abundance throughout the core.
The three zoom-in boxes allow to appreciate the almost perfect matching
of relative intensity and line profile of all the seven hyperfine
components that exhibit conspicuous excitation anomalies.
The double-peak profile of each features is produced by the slow
gravitational contraction motion.
See \cite{Bizzocchi:2013fk} for full details of the analysis. }
\label{fig:bizz}
\end{center}
\end{figure*}

Highly precise determination of transition rest-frequencies (obtained
when possible by sub-Doppler measurements \cite{Winton1970219}) are also important for
common and widely-used molecular tracers. These data make sophisticated
dynamical studies of star forming regions viable (see an example in
figure \ref{fig:bizz}).  Laboratory measurements are not limited to the main 
isotopic 
species and to vibrational ground state because a sound spectral knowledge of rare isotopologues and of 
vibrationally excited states allows to extract from observations important chemical and physical insight on the insterstellar gas, i.e., study isotopic fractionation mechanisms, estimate gas and dust temperature, and shed light on the complex, and 
presently poorly understood, gas-dust interactions.

In addition to the measurements accurate line positions and their analysis in order to provide accurate 
spectroscopic 
constants (that in turn can be used to accurately predict the rest frequencies of transitions not investigated in 
the 
lab), high-resolution spectroscopy can also efficiently support collisional dynamics studies. Indeed, in order to 
interpret observed interstellar spectra in terms of local physical conditions, one must consider the process of 
spectral 
line formation and this in turn requires a knowledge of rates for radiative and collisional excitation. 
Rotational 
state-to-state collisional rate coefficients are thus data of paramount importance for a proper modelling of line 
observations and represent another critical molecular physics' contribution to the astrochemistry.
The calculation of the collisional cross-sections and rate constants can be carried out in the close-coupling 
theoretical 
framework proposed by  \cite{Arthurs:1960zi} and implemented in the MOLSCAT code 
\cite{Hutson1994CoPhC..84....1H,Hutson:MOLSCAT}. These quantities 
are expressed in terms of the scattering matrix S which in turns, can be calculated by solving the time-
independent 
Schr\"odinger equation involving the interaction potential V of the colliding system. 	One thus ends with the 
problem of 
computing a high-accuracy potential energy surface (PES) for the interaction between the tracer molecule and a 
perturber 
(He or H$_2$), a task that can be efficiently tackled with ab initio theoretical chemistry methods. 

This procedure has also a recognised and fruitful link to laboratory studies. Using MOLSCAT, one may also derive 
the 
parameters describing the collisional broadening of the rotational transitions: the line-broadening and line-shift 
coefficients are related to the real and imaginary parts of the cross-section and of the efficiency function, both 
derived via the $S$ scattering matrix. 

Experimental information on the line-widths can then be obtained in the molecular spectroscopy laboratory as a 
natural 
side-products of the rest-frequency measurements carried out in controlled pressure and temperature conditions 
(pressure 
broadening measurements). Although these studies  can not directly estimate state-to-state rates, they provide an 
important experimental validation of the computational procedure. Parameters derived by line profile studies 
represent 
not only a stringent test of the accuracy of the PES used for the scattering calculations, but they also provide a 
mean 
to improve the theoretically computed potential by modifying it slightly (morphing) to fit the experimental data 
\cite{Meuwly:1999fk}.

\section{Perspective}

\subsection{The astronomical perspective}

The interest for the astrochemical modelers would be to find new pathways forming efficiently molecules at low 
temperature via reactions that were though to be negligible based on
high temperature measurements and assuming a purely Arrhenius behaviour at low-temperatures encounters in the ISM. 
The questions are how many of those kind of reactions exist and what would be their effects on gas-phase chemical 
models. Are ion-neutral reactions in the gas-phase and surface chemistry the only efficient ways to form 
molecules? Somehow we may have reach the �limit� of what
ion-neutral reactions can synthesis. Surface chemistry seems to be capable of making may observed species but 
there are so many parameters that one  can play to make it �match�. Neutral-neutral pathways are still lacking experimental/theoretical rates that are validated for mow temperature regimes, below some tens of Kelvin. 

Also, there is a need for state-to-state reaction rates with vibrationally excited \ce{H_2} (one way to go over the barriers).
In the low energy, the astrochemists are excited by the new experiments are very low temperatures where  somehow the rates for neutral-neutral are not low or even higher than the rates at $\sim 1000\mathrm K$. How
theory is coping with those results:(i) low temperature rates for the \ce{OH + CH_3OH} reaction \cite{Hickson:2016fk},; (ii) same, for the \ce{OH + H_2} reaction, \cite{2016JChPh.144q4303M}.

Collisional data are central in any quantitative interpretation of feature-rich astronomical spectra together with the availability of open-source sophisticated radiative transfer tools (e.g. RADEX, \citet{2007A&A...468..627V}, 
 LIME, \citet{2010A&A...523A..25B}). Astrochemical studies have focused in the past mostly on cold regions but recent (e.g., ALMA) and future instruments (like JWST) have/will have much improved sensitivities. Departures from local thermodynamical equilibrium (LTE) are dominant as lines from higher and higher levels are detected or when the interplay between the radiation from warm dust grains and that from the molecules is strong, the so-called infrared-pumped populations. The sub-millimeter array could observe vibrationally-excited rotational lines ($v=1$) of  CO lines toward the extreme carbon star IRC+10216. More recently the sub-millimeter ALMA telescope has the sensitivity to detect for asymptotic giant branch stars \cite{2016arXiv160803271K}.
Dedicated searches for $v=$1 lines not only of CO but also of large species like methanol will open an extra possibility to study warm molecular regions in addition to the $v=1-0$ transitions seen in the near- and mid-infrared in the future with the MIRI spectrometer on board the JWST. Indeed line surveys of template objects such as toward the class 0 protostellar binary IRAS 16293-2422 \cite{2016arXiv160708733J} or towards the Orion (see e.g.\citet{2010A&A...521L..21C}, for an example from the Herschel HEXOS program) show a large amount of unassigned lines that may be caused by rotational transitions within a vibrationally-excited level. 

Extreme cases of departure from Local Thermodynamical Equilibrium are seen in astronomical masers for many species 
(OH, CH, \ce{H_2O}, \ce{H_2CO}, \ce{CH_3OH}, SiO, HCN, \ce{NH_3}, SiS). The population 
inversion may occur when higher levels are populated by absorption of IR photons emitted by dust grains. Purely 
collisional masers are also observed and they are likely frequent in complex organic molecules, as demonstrated in 
the case of methyl formate. Such masing lines at radio wavelengths can be important for the identification of new 
complex species since definitive detection in the millimeter domain can prove challenging in line-rich sources. 
Masing lines are also found towards comets, planetary atmospheres, late-type star atmosphere, star-forming 
regions, and even extragalactic sources.

Therefore even for simple species such as CO, coverage of large number of rotational and vibrational levels  in theoretical works is needed. In addition to inelastic collisional data, reaction rates are needed for astrochemical modeling. Simultaneous modeling of the chemical composition and molecular excitation is becoming a routine endeavour.

This is numerically and computationally challenging and may require a re-thinking of the methods used so far in theoretical works.

\subsection{The mathematical and numerical perspectives}

In order to have a relevant impact on the reliability and precision on the chemical networks currently used (e.g. the UMIST, KIDA, or PDR networks), with their thousands of chemical rates,  it has been shown during the workshop that full quantum computation of chemical reactions remains today out of reach, because of the huge size of the problems (the handling of huge matrices) and probably, the algorithms employed.. 

Two ways out have been delineated, that try and circumvent both the full PES computation and the full quantum  computation: (i) large efforts have been devoted to the Transition state theory, which enables to compute at least rates for high enough temperatures, when quantum effects are not dominant. This remains today the bets --if not only-- choice for the many reactions that we have to deal with; (ii) for more precise large computations, time dependent methods, quantum or semi-classical have been evoked during the colloquium. While the former is used for systems comprising up to 4 or 5 atoms \cite{Zhang:2015xy}, the latter is not yet fully come of age 
\cite{biling-book,BILLING1984239,Babikov:2016fk}. Much more experience must be gained on this latter topic.

One method, not mentioned during the workshop, deserves a special mention, the ring polymer dynamics \cite{Craig2005JChPh.122h4106C,Zhang:2014qv,Suleimanov2013CoPhC.184..833S}, that mimics the quantum dynamics in the vicinity of the TST. Here also, ore experience must be gained in order to fully prove its ease of use, and range of precision.

It must be strongly underlined that the interaction between dynamical system theory in mathematics on the one hand, and TST or reaction rate theory remains unfortunately too tenuous. It is the the editor's opinion that much is to be gained if reaction rate theory would have taken full advantage of the counter-intuitive properties of classical dynamics near threshold (the TST). 

Most of the algorithms used in scattering insofar have been thought of and developed at a time when massively parallel computers
were unheard of. While the actual ab initio point wise computations seem not to be a bottleneck, all subsequent stages in the quantum dynamics are very time-consuming. There is no consensus, let alone universal practical method to fit economically a PES functional form  on a minimum number of ab initio points, especially for reactive scattering. No interactions with optimization techniques and mathematically oriented computational geometry have been described.

For most time-independent and time dependent quantum computations, most (up to 90--95\%) of the computer time is spent in linear algebra, even when the code is fully optimized and makes use of present-day linear algebra specialized packages, like BLAS and LAPACK. Progressing further relies on novel approches, like 'divide and conquer' strategies, or else on devising the problem with the parallelization in mind at the very beginning of the algorithms definitions and implementations.

In conclusion, \textbf{there is a real need for a large number of chemically oriented computations for astrophysics}, with the precision needed for relevant modeling. Many avenues remain open; they are ill-explored because  there had been of lack of interest in low-T, low density, and non equilibrium chemistry. Now that new observations, new experiments keep on bringing  results, theory, as a ways to model, understand, and predict observation and experiments must have the capabilities  to answer these new challenges.
 
\section*{Acknowledgment}
We thank the agencies providing means to our respective programs mentioned in the abstract: COST, MPG, MPIAA, CNRS.
The contribution from SJK is based on work supported by the U. S. Department of Energy, Office of Basic Energy Sciences, Division of Chemical Sciences, Geosciences, and Biosciences at Argonne under Contract No. DE-AC02-06CH11357. 
The SOC and the participants wish to thank A.~Langer (MPE Garching, Germany) and B.~Bishop (Open U., Milton 
Keynes, UK) for their  patient work.

\clearpage

\bibliographystyle{aipnum4-1}
\bibliography{white_paper.bib}

\begin{thebibliography}{127}
\bibitem [{Notea()}]{Notea}%
  \BibitemOpen
  \bibinfo {note} {Under normal conditions, 298K, 1 atmosphere, ideal gas has a
  density of $2.5\protect \tmspace +\thinmuskip {.1667em}10^{19}\protect
  \tmspace +\thinmuskip {.1667em}\protect \mathrm {cm^{-3}} $}\BibitemShut
  {NoStop}%
\end{thebibliography}
 
\clearpage

\sffamily
\begin{longtable*}{l@{\hspace{0.5ex}}p{6cm}@{\hspace{3ex}}l}
\multicolumn{3}{c}{ AUTHORS AND INTERVENTIONS DURING COLLOQUIUM AND WORKSHOP}\\ &  &\\
\hline
&  &\\
Speaker & Subject & Laboratory \\
& & \\\hline
& & \\
\emph{\textbf{Colloquium}}& &\\
& & \\

Paola Caselli & Introduction & MPE, Garching, Germany \\[3em]
Cecilia Ceccarelli & Molecular complexity in star forming regions: What we know and what we do not. & U. Grenoble, 
IPAG, France 
\\[3em]
Eric Herbst & Some Poorly Understood Classes of Interstellar Reactions & U. Virginia, USA \\[3em]
Wing-Fai Thi & Impact of collisional and chemical rates on astrochemistry modelling in protoplanetary disks & 
MPE, Garching, 
Germany \\[3em]
Nadia Balucani & Revisiting gas-phase chemistry in astrochemical models & U. Perugia, Italy \\[3em]
Liton Majumdar &
KIDA (Kinetic database for Astrochemistry): Present and Future & U. Bordeaux, France \\[3em]
Denis Duflot & Ab initio study of small molecules PES: 2 case studies & U. Lille, France \\[3em]
Jo\~ao Brand\~ao & Potential Energy Surfaces for reaction dynamics& U. Faro, Portugal \\[3em]
Alberto Rimola & 
Ab initio modeling of surface- induced reactions of astrochemical interest. Strategies and current limitations& 
U. Aut\`onoma 
Barcelona, Spain \\[3em]
Cristina Puzzarini & Rotational spectroscopy as a tool to investigate molecules in space&U. Bologna, Italy \\[3em]
Ad van der Avoird & From scattering resonances to rate coefficients for astrophysical modelling &U. Nijmegen  , 
The 
Netherlands \\[3em]
Stephen Klippenstein & A Priori Kinetics: Coupling Electronic Structure Theory with Statistics,
Dynamics, and the Master Equation & Argonne Nal. Lab. (IL) USA \\[3em]
Fran\c{c}ois Lique & Quantum scattering calculations: Closing the gap between
observations and astrochemical
models& U. Le Havre, France \\[3em]
Tijs Karman &Quantum mechanical calculation of the collision-induced absorption
spectra of N$_2$ and O$_2$ & U. Nijmegen  , 
The 
Netherlands \\[3em]
Tamiki Komatsuzaki & Phase space geometry and chemical reaction dynamics: Past, present and future.
 & U. Sapporo, Japan \\[3em]
Mikito Toda & Dynamical reaction theory: Beyond the conventional statistical reaction theory& Nara Women U., Japan 
\\[3em]
Holger Waalkens & Quantum Transition State Theory & U. Groningen, The Netherlands \\[3em]
Alexandre Faure & Classical approach to energy transfer: Successes and limitations & U. Grenoble , France \\[3em]
Hiroshi Teramoto & Dynamical Reaction Theory: Beyond  the conventional perturbation theory & U. Sapporo, 
Japan \\[3em]
Dmitry Babikov & Recent advances in the mixed quantum/classical theory of inelastic
scattering & Marquette U., Milwaukee, USA \\[3em]
Fabien Gatti & MCTDH & U. Montpellier, France \\[3em]
& & \\
\hline
& & \\
\emph{\textbf{Workshop}}& &\\
& & \\[3em]
 Tamiki Komatsuzaki & Status of the mathematical TST (1) &  \\
  & Networks as dynamical systems & U. Sapporo, Japan  \\[3em]
   Eric Herbt & Current problems in Astrochemistry & U. Sapporo, Japan \\[3em]
 Jo\~ao Brand\~ao & Ab initio status, PES for open-shell systems & U. Algarve, Portugal\\[3em]
 Holger Waalkens & Status of the mathematical TST, roaming, dynamical systems  
  \&~Quantum TS &  U. Groningen, The Netherlands \\[3em] 
 Stephen Klippenstein & Transition State Theory, radical-radical reactions & Argonne Nal. Lab. (IL), USA \\[3em]
 Ad van der Avoird & Zeeman spectroscopy; inelastic collisions &  U.Nijmegen, The Netherlands \\[3em]
 Octavio Roncero & H$_n^+$ systems & CSIC, Madrid, Spain \\[3em]
 Dmitris Skouteris & Time-dependant and time-independant reaction quantum dynamics &  Scuole Normale Superiore, 
Pisa, 
Italy \\[3em]
  Mikito Toda & Crises in chaotic scattering & Women U. Nara, Japan\\[3em]
 Hiroshi Teramoto  & Bifurcations in NHIMs & U. Sapporo, Japan \\
\end{longtable*}

 \end{document}